\documentclass[
superscriptaddress,
reprint,
 amsmath,amssymb,
 aps,
]{revtex4-2}

\usepackage{graphicx}
\usepackage{dcolumn}
\usepackage{bm}
\usepackage{xcolor}
\usepackage{comment}
\usepackage[normalem]{ulem}
\usepackage[T1]{fontenc} 
\usepackage{epsfig}
\usepackage{comment}

\usepackage{xspace} 
\newcommand\capa{({\bf a})\xspace} 
\newcommand\capb{({\bf b})\xspace} 
\newcommand\capc{({\bf c})\xspace} 
\newcommand\capd{({\bf d})\xspace} 
\newcommand\cape{({\bf e})\xspace} 

\newcommand{\kex}{\kappa_\mathrm{ex}}
\newcommand{\ki}{\kappa_\mathrm{i}}

\usepackage[draft,inline,nomargin,index]{fixme} 

\fxsetup{theme=color,mode=multiuser}

\FXRegisterAuthor{Ofer}{aOfer}{\color{brown} Ofer} 
\FXRegisterAuthor{omri}{aOmri}{\color{red}  Omri}
\FXRegisterAuthor{barak}{aBarak}{\color{blue}  Barak}
\FXRegisterAuthor{yair}{aYair}{\color{blue}  Yair}
\FXRegisterAuthor{todo}{aTodo}{\color{red}  Todo}

\begin{document}

\preprint{APS/123-QED}

\title{Single-atom trapping in the evanescent field of an integrated photonic resonator}

\author{Yair Margalit}\thanks{These authors contributed equally to this work.}
\author{Omri Davidson}\thanks{These authors contributed equally to this work.}
\author{Oded Zemer}%
\author{Yoad Michael}
\author{Orel Bechler}
\author{Dror Liran}
\author{Noam Gross}
\author{Doron Azoury}
\author{Jeremy Raskop}
\author{Yaakov Yudkin}
\author{Gabriel Guendelman}
\author{Moshe Katzman}
\author{Michael Nagli}
\author{Yair Antman}
\author{Nadav Kandel}
\author{Geva Arwas}
\author{Idit Peer}
\affiliation{Quantum Source Labs, Ilan Ramon 3 St., Ness Ziona, 7403636, Israel
}%

\author{Ofer Firstenberg}

\affiliation{Department of Physics of Complex Systems, Weizmann Institute of Science, Rehovot, 76100, Israel}%

\author{Barak Dayan}
 \email{barak.dayan@weizmann.ac.il}
\affiliation{Quantum Source Labs, Ilan Ramon 3 St., Ness Ziona, 7403636, Israel
}
\affiliation{AMOS and Department of Chemical Physics, Weizmann Institute of Science, Rehovot, Israel.}

\date{\today}

\begin{abstract}
Strong atom-photon interactions on scalable photonic platforms
hold significant potential for both atomic and photonic quantum information platforms. In particular, trapping of a single atom on a planar photonic integrated resonator at the subwavelength distances required for strong coupling to the guided modes has remained an outstanding challenge. Here we demonstrate efficient trapping of a single ultracold rubidium atom within the evanescent field of an integrated silicon-nitride microring resonator, at distances of $150\text{--}200$ nm from the chip surface.
Efficient, single-stroke loading process is achieved using an evanescent-field mechanism related to Sisyphus cooling, in which a single scattering event dissipates the atom's kinetic energy and transfers it into a near-surface trap. We observe logarithmic scaling of trapping durations spanning from sub-millisecond timescales up to 1 second, without continuous cooling. The trapped atom couples efficiently to the resonator, enabling on-chip photon collection, photon antibunching, and Purcell-enhanced spontaneous emission with single-atom cooperativity exceeding unity. Our results establish the potential of CMOS-compatible chip-based atom-photon interfaces for scalable quantum photonic circuits.
\end{abstract}

\maketitle

\section{\label{sec:level1} Introduction}

Single atoms coupled to photonic integrated circuits (PICs) offer a powerful route towards scalable quantum technologies, combining the long coherence times of atomic systems with the compactness, functionality, stability, and scalability of integrated photonics \cite{OBrien-2009, Chang-2018}. Coupling single atoms to single guided modes opens the path to deterministic single-photon nonlinearity  \cite{PhysRevA.73.062305, PhysRevLett.133.113601, Chang-2014, Firstenberg-2017}, vital for non-reciprocal operations and photon-atom quantum gates \cite{Duan.Kimble.PRL.2004, Photon.Turnstile.Science.2008, Reiserer2014AtomPhotonGate, Shomroni-2014, Volz2014NonlinearPhaseShift, Tiecke-2014, Rosenblum-2016, Scheucher2016QuantumCirculator, Lodahl-2017, Bechler-2018} - capabilities considered pivotal for photonic quantum computation \cite{OBrien-2009, Lodahl-2015, Aqua-2025} and quantum networks \cite{Kimble-2008, Ritter2012ElementaryNetwork, Wehner-2018}. 

Trapping of individual single atoms has been demonstrated next to one-dimensional nano-photonic platforms, such as suspended photonic band-gap structures \cite{Thompson-2013, Menon-2024} and whispering-gallery-mode microresonators \cite{Will-2021}. Additionally, arrays of single-atom traps have been implemented next to suspended photonic band-gap structures \cite{Goban-2014, Goban-2015, Menon-2024} and in the evanescent field of nano-fibers \cite{Arno.PhysRevLett.104.203603,Corzo-2016, Corzo2019WaveguideCoupledSingleExcitation, Arno2023.PRXQuantum.4.040308, pennetta2025hybridtrappingcoldatoms}. 
However, the scalability and mechanical stability of PICs have motivated an effort towards trapping single atoms next to their surface in a way that enables efficient coupling to the integrated single-mode resonators and waveguides. Single atoms were trapped next to a PIC waveguide \cite{Kim-2019}, and atomic ensembles were trapped on a photonic integrated resonator, achieving long lifetimes at distances of several hundred nanometers from the surface \cite{Hung-2024}. Transient single-atom interaction with a photonic integrated resonator was observed via atom guiding without trapping \cite{Hung-2023}.   
These demonstrations highlight the fundamental challenge in achieving stable trapping of a single atom that couples efficiently to stable and scalable systems such as PIC-based resonators.

\begin{figure*}
\centering
\includegraphics[width=0.97\textwidth]{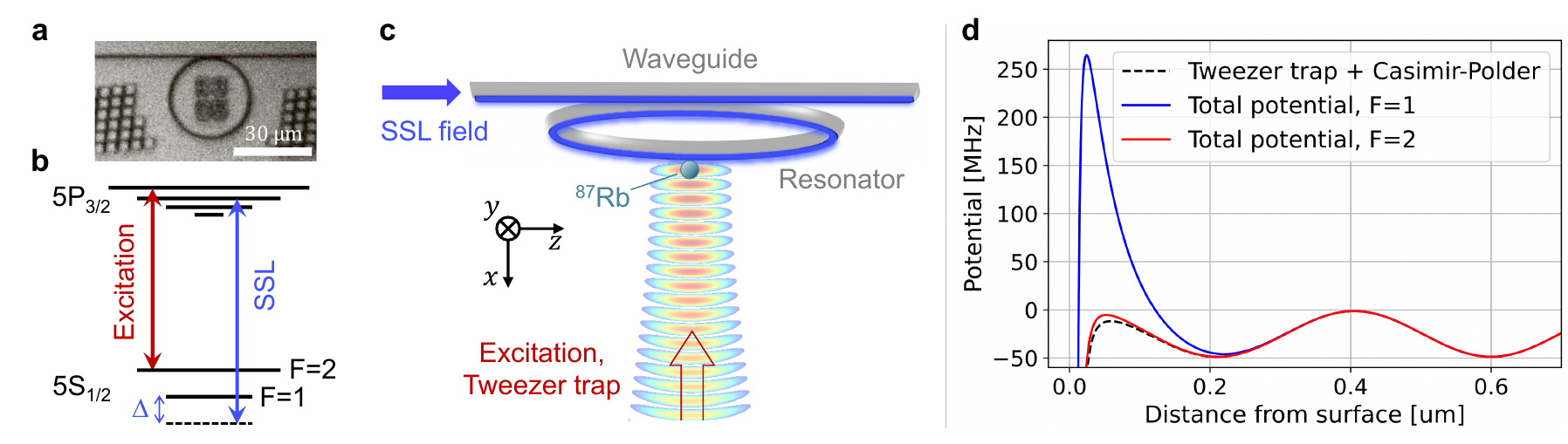}
\caption{\textbf{Schematic of the experiment.} \capa Optical image of the chip showing the SiN resonator and bus waveguide.
\capb Atomic energy-level diagram and optical fields (not to scale). The Single-Stroke-Loading (SSL) detuning $0\le\Delta\le300$\,MHz is much smaller than the ground-state hyperfine splitting of 6.8\,GHz.
\capc Atoms are launched upwards from a magneto-optical trap towards the chip, where they experience 
the evanescent SSL field (blue),
which decelerates them and loads them into a near-surface standing-wave trapping potential formed by a tweezer beam reflected from the chip.
\capd Atomic ground-state energy levels as a function of distance from the chip surface. Dashed line shows the tweezer trapping potential combined with the surface-induced Casimir-Polder. Red and blue lines include the addition of the single-stroke loading field for $F=2,1$, respectively. The atoms are initially prepared in the $F=1$ ground state, approach the chip, and decelerate in the conservative potential formed by the evanescent field of the light in the resonator. A single photon scattering event can transfer the atom to the $F=2$ state, which experiences a trapping potential. } 
\label{fig:schematic}
\end{figure*}

Here we demonstrate, for the first time, loading and stable trapping of a single ultracold rubidium atom within the evanescent field of a planar, chip-based SiN microring resonator, using both an evanescent field and a reflected dipole trap. Loading into the trap is enabled by an optical 'crash-cushion' composed of a near-resonant blue-detuned evanescent-field repelling potential, which slows down the approaching atoms, until a photon scattering event transfers them into a further-detuned hyperfine manifold (similar to evanescent-wave cooling in surface traps \cite{Soding-1995, Ovchinnikov-1997, Hammes-2002}), which experiences mainly the trapping potential formed by a reflected dipole trap. This single-stroke loading (SSL) process  captures individual atoms at distances of $150\text{--}200$\,nm from the chip surface, even without continuous laser cooling
(which typically requires unrestricted optical access, available only in
suspended nano-structures), or sub-microsecond active feedback. This approach is also largely insensitive to the initial velocity of the atom and does not require its kinetic energy to be lower than the trap depth (in contrast to the 'sudden turn-on' mechanism \cite{Will-2021}).
Once trapped, the atom couples efficiently to the guided mode of the resonator, allowing fluorescence photons to be collected and routed on chip. Single-atom occupancy is verified through pronounced photon antibunching, while time-resolved pulsed excitation reveals Purcell-enhanced spontaneous emission and single-atom cooperativity $C>1$. The trapped atoms exhibit dynamics likely governed by near-surface tunneling, with logarithmic distribution of trapping durations spanning from sub-millisecond timescales to one second. These results present the potential of this chip-integrated atom-photon platform as a scalable route towards PIC-based quantum nonlinear optics.

\section{Setup}

Our system incorporates a small ultra-high-vacuum (UHV) cell, where the chip, facing down, is glued to seal a hole at the top of the cell. The PIC is fabricated using a commercial low-loss low-pressure chemical vapor deposition (LPCVD) process that yields a 400~nm silicon-nitride (SiN) layer on top of a $3\,\mu$m silica substrate, which covers the silicon wafer. The PIC includes a $30\,\mu$m diameter SiN ring resonator supporting a transverse-magnetic (TM) mode at $780$~nm, near critical coupling to a bus waveguide with external and intrinsic field coupling rates  $\kex\approx \kappa_\mathrm{i}=1.16\pm0.02$~GHz [Figs.~\ref{fig:schematic}a,c; see Sec.~\ref{SM rsonator fit} of the supplementary material (SM) for extraction of the resonator parameters].
Both the resonator and the waveguide are $450$~nm wide.
With a free spectral range of $\mathrm{FSR}=1.36$~THz, the corresponding finesse is $\mathcal{F}=\mathrm{FSR}/(2\kappa) \simeq 300$, where $\kappa=\kex+\ki$. Optical excitation of, and readout from, the resonator are performed using lensed fibers coupled to the bus waveguide with efficiencies of 50\%.

The resonator-mode evanescent field extends to the vacuum region below the chip and decays as $\exp(-x/\Lambda)$, where $x$ is the distance from the surface, and $\Lambda = 86$~nm, evaluated numerically (see Sec.~\ref{trap and g simulations} of the SM). This evanescent field plays a dual role in our system. First, it enables Purcell-enhanced efficient coupling of photons from the trapped atom to the resonator and the waveguide, which are subsequently detected by superconducting nanowire single-photon detectors (SNSPDs). Second, it supports the blue-detuned 'crash-cushion' SSL field, which dissipates the kinetic energy of the launched atom, 
allowing its efficient loading into a dipole trap formed by a tightly focused, red-detuned optical tweezer beam that is introduced through an NA=0.28 objective lens below the cell. The tweezer beam is partially reflected from the chip surface, forming a  standing-wave potential that provides both vertical and transverse confinement in the vicinity of the resonator, as illustrated in Fig.~\ref{fig:schematic}c. The tweezer beam has a $1/e^2$ waist radius of 1.5~$\mu$m and operates at wavelengths ranging between 820 and 850~nm. As shown in Fig.~\ref{fig:schematic}d, the total potential experienced by an atom near the chip surface is given by the sum of the reflected optical tweezer potential, the Casimir-Polder potential from the chip, and the state-dependent potential of the evanescent SSL field.   
The position of the dipole potential minimum closest to the surface depends on the structure and thickness of the dielectric layers constituting the PIC, as well as on the tweezer wavelength which allows to tune the trapping distance from the surface (see Sec.~\ref{trap and g simulations} of the SM). For the tweezer wavelengths used here the resulting trap minimum was varied between 150 and 200\,nm from the resonator surface. At the closest potential minimum the single-photon coupling rate between the atomic dipole and the resonator mode is 
$g \sim 100$MHz, resulting in single-atom cooperativity $C= g^2 /(\kappa \gamma)\sim 1$, where $2\gamma=6.06$ MHz is the atomic spontaneous emission rate to free space (see Sec.\,\ref{sec:characterization}). Atoms are excited using the $F=2\rightarrow F'=3$ cycling transition, using the same optical mode as the tweezer beam.

For the trap powers employed in our experiment the vertical trap frequencies are on the order of $500\,$kHz, with corresponding trap depths reaching up to $\sim 100$\,MHz ($\sim 5$\,mK). The dependence of the trap position, depth, and motional frequencies on the tweezer wavelength and power is characterized experimentally and discussed in detail below in the Results, Methods, and SM sections.

\section{Experimental sequence and Single-stroke loading}
Our experimental sequence begins by cooling $^{87}$Rb atoms in a magneto-optical trap (MOT) to a temperature of $30\,\mu$K, and subsequently launching them upwards using a moving-molasses sequence towards the resonator (see Methods for details). 
The resonator is frequency-locked to the $F=2\rightarrow F'=3$ cycling transition of the $^{87}$Rb D2 line at 780~nm (Fig.~1b). 

Atoms approaching the surface are initially prepared in the $F=1$ ground state, which experiences a repulsive potential generated by the evanescent SSL field (Fig.~\ref{fig:schematic}d, blue line). The blue detuning $\Delta$ of the SSL field from the $F=1\rightarrow F'=2$ D2 transition is varied in our system from zero up to $300~$MHz. As the atoms climb the optical potential barrier, they lose kinetic energy. For an appropriate choice of detuning and intensity of the loading field, an approaching atom is likely to undergo a single photon-scattering event that may transfer it to the $F=2$ ground state (according to the atomic branching ratio). For this state, the repulsive potential is drastically lower, as it is $\sim6.8$\,GHz further detuned, resulting in a considerable loss of energy, thousands of times larger than the single-photon recoil energy. Since the optical potential is now dominated by the near-surface optical tweezer field (Fig.~\ref{fig:schematic}d, red line), the atom is likely to remain within the trapping region. Crucially, a single, irreversible scattering event suffices to load the atom with high probability, in a manner that is relatively insensitive to the incoming kinetic energy (see Sec.~\ref{Loading simulation} of the SM).

Evanescent-wave schemes have previously been used to cool ensembles of atoms near dielectric surfaces, typically at distances of tens of micrometers, exploiting differential light-shifts between hyperfine states to dissipate kinetic energy through controlled photon scattering~\cite{Ovchinnikov-1997, Hammes-2002}. In these gravito-optical surface traps (GOST) and related implementations, the confinement is relatively weak and the scattering probability per interaction with the evanescent field is intentionally kept low
($\sim 1\%$), to minimize heating and atom loss, allowing many cooling cycles. 
Related concepts have also been explored for continuous loading of atoms from a beam using optical fields~\cite{Pfau-2014}, and single-photon cooling has also been demonstrated for ensembles in free-space\,\cite{Price-2008}.

In contrast, here we operate in a regime optimized for efficient loading of individual atoms into a tightly confined near-surface trap. In our geometry, atoms are launched upwards toward the chip surface in a fountain configuration and therefore have only a single opportunity to be captured. This fundamentally changes the optimal operating point: rather than minimizing scattering, we seek to maximize the probability of a single irreversible Raman transfer into the trapped state. We achieve this by balancing the intensity and detuning of the evanescent loading field with respect to the lower hyperfine state, reaching trapping probability per SSL pulse of the order of $30\%$, with variations depending on the atomic flux. In this regime, the dynamics are dominated by a single scattering event that transfers the atom into the trapping potential, rather than by repeated scattering cycles that gradually cool the atomic motion.

Figure 2 summarizes the experimental results of the evanescent SSL process. Following preparation of the atoms in the $F=1$ ground state and the arrival of the atomic cloud at the PIC, a sequence of 0.5~ms SSL pulse followed by a 2~ms excitation pulse is applied. If an atom is successfully captured during the loading pulse, it subsequently scatters photons from the following excitation pulse, which is near-resonant with the cycling transition of the trapped atom (see Methods), until it escapes the trap. The photons scattered into the resonator are collected through the waveguide and detected by the SNSPDs.

\begin{figure*}
\centering
\includegraphics[width=1\textwidth]{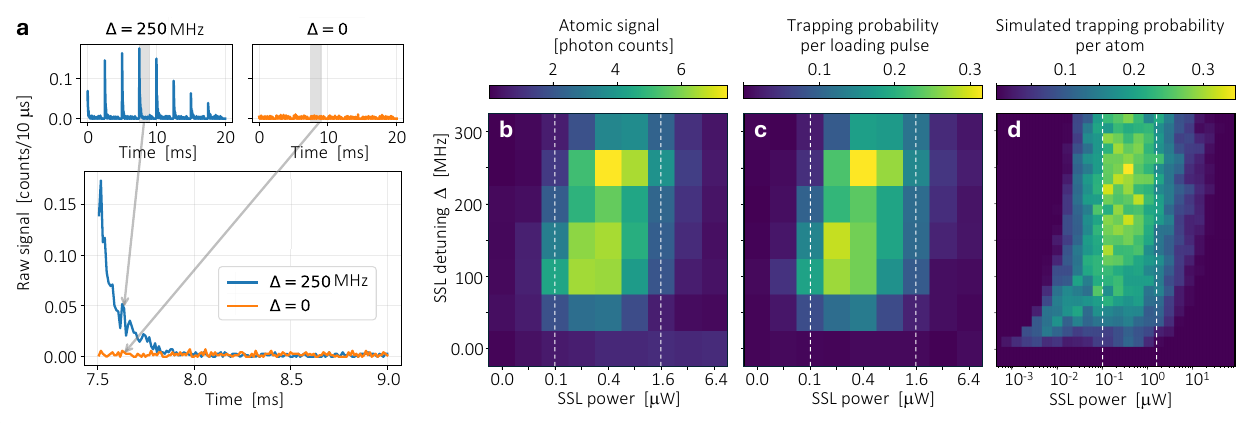}
\caption{\textbf{Atom trapping via evanescent single-stroke loading.} \capa Atom trapping signal. \textbf{Bottom:} raw photon histogram counts collected from the PIC when the excitation field is applied following an evanescent-field loading pulse, which is either with positive detuning $\Delta$ (blue) or on-resonance (orange) relative to the bare atomic transition. For positive detuning, the photon counts exhibit a pronounced decaying signal at the beginning of the excitation window, indicating fluorescence from a trapped atom that subsequently escapes the trap. In contrast, for the on-resonance loading field, the photon counts remain constant in time, only marginally exceeding the background level due to imperfect optical pumping into the $F=1$ ground state. \textbf{Top:} raw photon histograms for multiple repetitions of the loading and excitation sequence, shown for (left) blue-detuned and (right) on-resonance evanescent fields. Photon counts are not recorded during the loading pulse. The envelope of enhanced counts at the start of each excitation window reflects the time-dependent atomic flux arriving at the PIC during the MOT cloud transit. 
\capb Total detected atomic signal per atomic cloud transit as a function of evanescent-field detuning and input power to the PIC (logarithmic power scale for non-zero power). 
\capc Trapping probability at the peak of the atomic cloud transit as a function of evanescent-field detuning and power. The close correspondence between the maps in (b) and (c) indicates that the detected signal is dominated by fluorescence from trapped atoms.
\capd Simulated trapping probability as a function of evanescent-field detuning and intensity, showing good agreement between experiment and simulation for the required power and detuning yielding peak loading efficiency. The SSL power range of $0.1-1.6\,\mu W$, in which the maximal atom trapping was measured, is marked by dashed white lines in \capb -\capd. }
\label{fig:param_scan}
\end{figure*}

Figure~\ref{fig:param_scan}a shows the raw photon histogram counts recorded during the excitation pulse following an evanescent SSL pulse that is either blue detuned or on resonance with respect to the bare $F=1\rightarrow F'=2$ atomic transition. For positive detuning, a pronounced fluorescence signal is observed at the beginning of each excitation window and decays on a timescale of $\sim 0.1$ ms, consistent with emission from a trapped atom that is subsequently lost from the trap. The overall envelope of the signal (top-left panel) reflects the time-dependent atomic density arriving at the chip during the MOT cloud transit. In contrast, no corresponding signal is observed for the on-resonance SSL field, confirming that trapping occurs only in the evanescent SSL regime.

We integrate the SNSPD counts over each atomic cloud transit to extract the total detected atomic signal as a function of evanescent loading-field detuning and power, as shown in Fig.~\ref{fig:param_scan}b. A pronounced atomic signal appears only for positive detunings and within a finite range of powers. This behavior reflects the interplay between the energy-conservative deceleration by the evanescent optical potential and the probability for a scattering event that may transfer the atom into the $F=2$ trapped internal state. Achieving the required combination of sufficient deceleration and photon scattering probability requires increasing power and detuning together; otherwise the scattering may occur before the atom is decelerated enough, or the atom may bounce back or hit the surface before scattering to the $F=2$ ground state (see Fig.\,S2 discussion in the SM).

In the following analysis we define a trapped atom event when at least two photons are detected within the first 500~$\mu$s of the excitation pulse. Figure~\ref{fig:param_scan}c shows the corresponding trapping-probability map evaluated at the peak density of the atomic cloud. The map closely mirrors the integrated atomic signal in Fig.~\ref{fig:param_scan}b, demonstrating that the detected signal originates predominantly from trapped atoms rather than transient scattering. The trapping probability reaches a maximum of 32\% per SSL pulse at a detuning of 250~MHz relative to the bare atomic transition and at an input power of 400~nW at the fiber coupled to the PIC (corresponding to $\sim2\,\mu$W circulating in the resonator, or a saturation parameter of approximately $4\times10^5$ of the evanescent field on the waveguide surface). The probability of falsely identifying a trapped atom is low, $0.3\pm0.2\%$, estimated from the probability to detect two photons at the last 500~$\mu$s of the excitation pulse (at which no atoms are expected, see \ref{fig:param_scan}a).

To gain insight into the loading mechanism, we perform a one-dimensional semi-classical simulations of the atomic dynamics. The position $x$ and momentum $p$ of the atom are treated classically, while its internal degrees of freedom are the occupation of the two atomic ground states $F=1,2$, which can change according to the local interaction with the evanescent loading field. In turn, the internal state determines the optical potential experienced by the atom and feeds back on its motional dynamics. Figure~\ref{fig:param_scan}d shows the simulated trapping probability for a single atom initially prepared in the $F=1$ state and approaching the surface with an incoming velocity of 0.3\,m/s. The simulation is in good agreement with the experimentally observed loading probability, including the optimal detuning and a limited optimal range for intensity. Further details of the model and simulations are provided in Sec.~\ref{Loading simulation} of the SM.

\section{Trapped atom characterization} \label{sec:characterization}

\begin{figure}
\centering
\includegraphics[width=0.5\textwidth, angle=0]{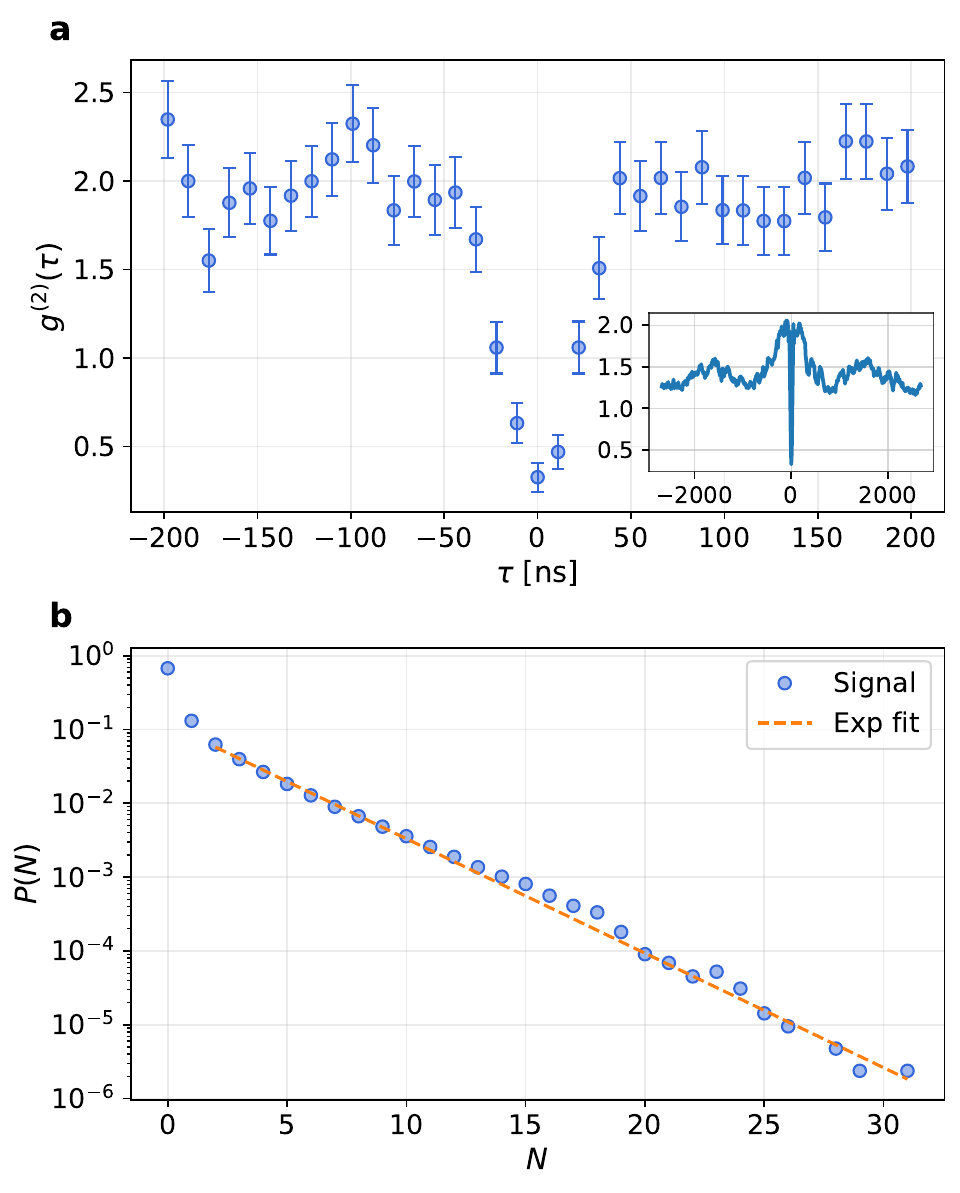}
\caption{\textbf{Photon statistics from a trapped atom.} \capa Single-atom verification via photon antibunching. Normalized intensity autocorrelation function $g^{(2)}(\tau)$ of photons emitted into the resonator, conditioned on detection of a trapped atom. 
A clear antibunching dip of $g^{(2)}(0)=0.33\pm0.08$ is observed at zero delay, confirming single-atom occupancy of the trap.
The antibunching dip is contrasted by the bunching of $g^{(2)}\approx2$ at $\sim 50\text{--}250$\,ns, which arises from the oscillations of the atom in the trap in the $x$ direction (see inset). These oscillations modulate the atom-resonator coupling efficiency at a frequency of approximately 635 kHz
(see Methods for the
independently measured trap frequencies).
\capb Probability $P(N)$ to detect $N$ photons within a 500~$\mu$s window from the start of the excitation pulse. For a trapped atom (conditioned on $N\ge 2$), the photon number distribution exhibits an exponential decay, as highlighted by the logarithmic vertical axis and the exponential fit. This behavior is consistent with variations in the loading depth achieved during the evanescent single-stroke loading process, due to a broad distribution of initial trapping conditions.
}
\label{fig:g2}
\end{figure}

We begin the characterization of the trapped atom by measuring the normalized second-order intensity autocorrelation function $g^{(2)}(\tau)$ of photons emitted into the resonator, where $\tau$ denotes the time delay between detection events. The correlations are performed between the two SNSPDs located at opposite outputs of the PIC. Classical light fields satisfy $g^{(2)}(0) \ge 1$, whereas antibunching with $g^{(2)}(0)<0.5$ is a hallmark of single-photon emission and indicates coupling to a single quantum emitter.

Figure~\ref{fig:g2}a shows the measured $g^{(2)}(\tau)$ conditioned on the detection of a trapped atom, as defined above (see Sec.~\ref{Data analysis SM} of the SM for the details of the $g^{(2)}(\tau)$ analysis). While timescales above $50\,$ns exhibit bunching up to $g^{(2)}\approx 2$ due to the approximately $635\,$ kHz oscillations of the atom in the trap (see Methods for the independently measured trap frequencies), a pronounced sub-Poissonian, antibunching dip with $g^{(2)}(0)=0.33\pm0.08$ is observed, demonstrating that there is a single trapped atom coupled to the resonator. 

Figure~\ref{fig:g2}b shows the probability $P(N)$ to detect $N$ photons within the first 500\,$\mu$s of the excitation pulse. Conditioned on a trapped atom ($N\ge 2$), the photon-number distribution exhibits an approximately exponential decay. 
We attribute these non-Poissonian statistics to variations in the initial trapping conditions established during the SSL process: as atoms can be loaded at different energies in the trapping potential, and since photon emission is accompanied by heating, deeper trapping allows an atom to scatter more photons before being lost. The varying trapping conditions are also manifested in a wide range of trapping durations without excitation (see Fig.\,\ref{fig:lifetime}).

Having established single-atom trapping and verified single-emitter behavior, we next characterize the coupling between a trapped atom and the photonic resonator. Figure~4a shows time-resolved fluorescence measurements performed by exciting a trapped atom with 10\,ns long optical pulses and recording the subsequent emission into the resonator mode. The detected fluorescence exhibits a clear exponential decay following the excitation pulse, indicating spontaneous emission into the guided mode.

From an exponential fit to the decay, we extract an excited-state $1/e$ lifetime of $(2\Gamma)^{-1}=16.3\pm0.4$~ns, significantly shorter than the free-space lifetime of rubidium, $(2\gamma)^{-1}=26.2$~ns \cite{Volz-1996}. This reduction reflects Purcell-enhanced spontaneous emission into the resonator with a single-atom cooperativity of $C=\Gamma/\gamma-1=0.61\pm0.04$ for a tweezer wavelength of $\lambda_\mathrm{t}=835$ nm. 

\begin{figure}
\centering
\includegraphics[width=0.5\textwidth]{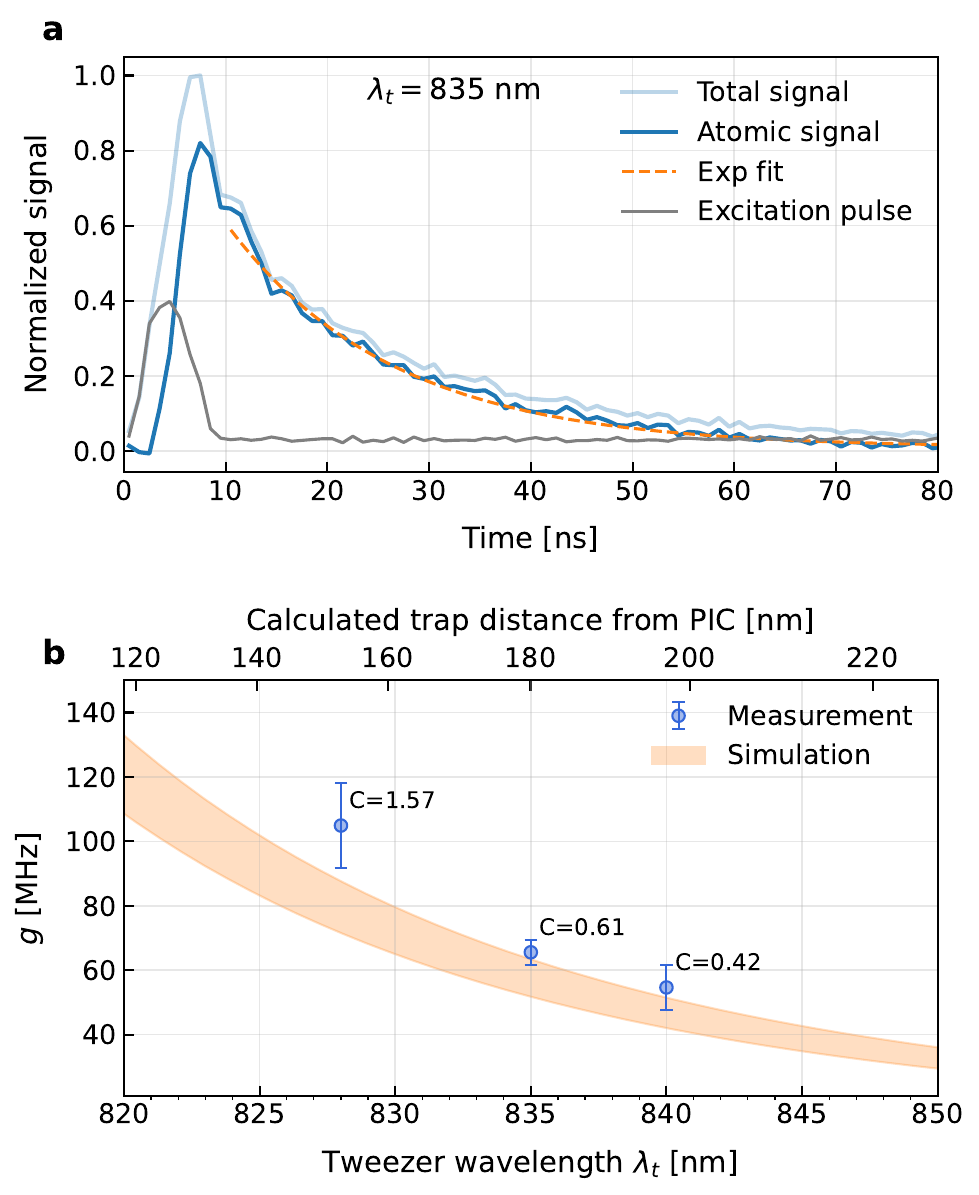}
\caption{\textbf{Atom-resonator coupling and Purcell enhancement.} \capa Time-resolved fluorescence from a trapped atom. Atoms are excited using short optical pulses (gray), and the subsequent fluorescence emitted into the resonator is detected (blue; obtained by subtracting the leaked excitation pulse from the total signal). An exponential fit to the decay following the excitation pulse (dashed orange) yields a Purcell-enhanced excited-state lifetime of $16.3\pm0.4$ ns, corresponding to a single-atom cooperativity of $C=0.61\pm0.04$ for a tweezer wavelength of $\lambda_\mathrm{t}=835$ nm. \capb Atom-resonator coupling strength versus trap position. Measured coupling rate $g$ extracted from the cooperativity $C=g^2/(\kappa\gamma)$ as a function of the tweezer wavelength $\lambda_\mathrm{t}$. The top horizontal axis indicates the distance of the trap minimum from the PIC for different $\lambda_\mathrm{t}=828, 835, 840$\,nm, as calculated by a numerical simulation that takes into account the multiple dielectric layers of the PIC (see Sec.~\ref{trap and g simulations} of the SM). The shaded region shows the simulated coupling rate with no fit parameters, where its width reflects the uncertainty associated with the unknown distribution of atomic Zeeman sublevels (see Sec.~\ref{Data analysis SM} of the SM). The measured coupling rates exceed the value calculated by the static simulation, which we attribute to atomic motion in the trap that enhances photon collection when the atom oscillates closer to the resonator surface (see Fig.~\ref{fig:g2}a).
}
\label{fig:cooperativity}
\end{figure}

\begin{figure}
\centering
\includegraphics[width=0.5\textwidth]{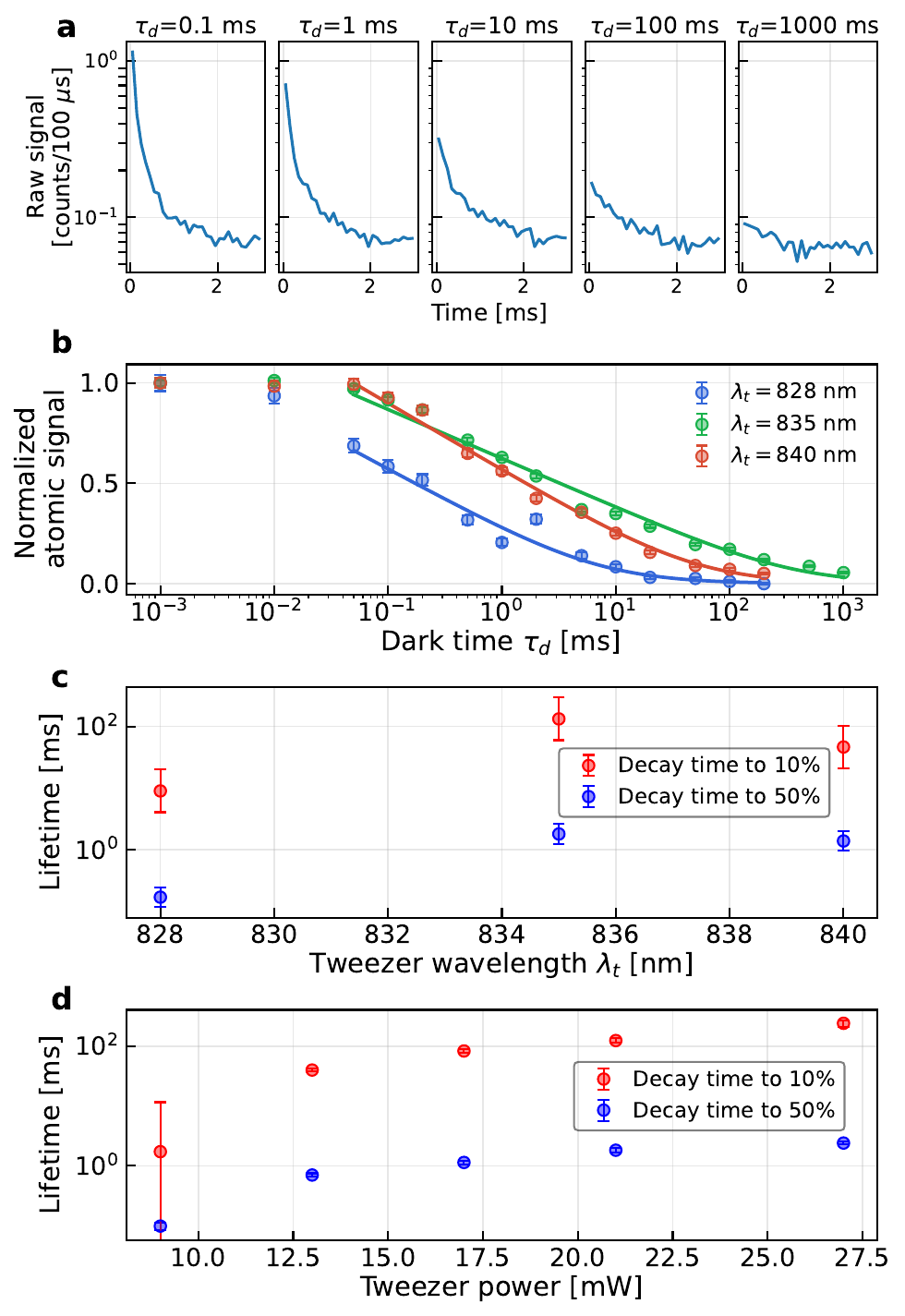}
\caption{\textbf{Lifetime of a trapped atom near the photonic chip.}
\capa Time-resolved survival signal.
Time histograms of the raw photon counts for different dark delays $\tau_\mathrm{d}$ between the end of the evanescent single-stroke loading pulse and the start of the excitation pulse. The vertical axis is shown on a logarithmic scale. 
\capb Normalized atomic signal versus dark time.
Total photon counts, normalized to the signal at $\tau_\mathrm{d}=0$, as a function of $\tau_\mathrm{d}$ for different tweezer wavelengths $\lambda_\mathrm{t}$.
For $\tau_\mathrm{d} \gtrsim 50~\mu$s, the signal decays approximately logarithmically with $\tau_\mathrm{d}$, as indicated by the logarithmic horizontal axis and the fits (solid lines). The tweezer power is 20 mW for $\lambda_\mathrm{t}=828$ and 840 nm, and 27 mW for $\lambda_\mathrm{t}=835$ nm.
\capc Trap lifetime versus trap position.
Lifetime, extracted as the dark time at which the atomic signal decays to 50\% (blue) and 10\% (red) of its initial value, plotted as a function of the tweezer wavelength $\lambda_\mathrm{t}$. The tweezer power is fixed at 20 mW.
\capd Trap lifetime versus tweezer trap power.
Lifetime extracted as in (c), shown as a function of the tweezer power for $\lambda_\mathrm{t}=835$ nm. Error bars in (c) and (d) represent uncertainties extracted from repeated measurements and single-measurement decay fits, respectively (see Sec.~\ref{Data analysis SM} of the SM). 
}
\label{fig:lifetime}
\end{figure}

To further quantify the atom-resonator interaction, we extract the coupling rate $g=\sqrt{C\kappa\gamma}$ for different trap positions by varying the tweezer wavelength, which shifts the location of the trap minimum relative to the chip surface. Figure~\ref{fig:cooperativity}b shows the measured coupling strength $g$ over the range $\lambda_\mathrm{t}=828\text{--}840$ nm, together with the corresponding atom-surface distances of $150\text{--}200$ nm, defined by the position of the trap minimum in the calculated optical potential. As expected, the coupling strength increases as the atom is trapped closer to the resonator, reflecting the exponential spatial profile of the evanescent field. For the smallest trap distance, obtained at $\lambda_\mathrm{t}=828$ nm, we measure a cooperativity of $C=1.57\pm0.36$, corresponding to a coupling rate of $g=105\pm13$ MHz.

The measured values of $g$ are compared with calculations for an atom located at the trap minimum, shown by the shaded region in Fig.~\ref{fig:cooperativity}b. The shaded range reflects the uncertainty associated with the internal Zeeman-state distribution of the trapped atoms. The upper boundary corresponds to maximally polarized atoms coupled via the cycling transition $|F=2,\ m_F=2 \rangle \rightarrow |F'=3,\ m_F'=3\rangle $, which yields the largest possible value of $g$ for the given resonator mode (see Sec.~\ref{Data analysis SM} of the SM for details). The experimentally extracted values of $g$ consistently exceed the static simulation. We attribute this discrepancy to the oscillatory motion of the atom in the trap, also observed in the intensity autocorrelation measurements (Fig.~\ref{fig:g2}a), which increases photon collection when the atom oscillates closer to the resonator surface and thereby biases the measured decay rate toward larger values.
Overall, the data support the assignment of the trapping positions and affirm that atoms are trapped at distances down to 150\,nm from the photonic integrated resonator.

Finally, we characterize the lifetime of atoms in the near-surface trap. To this end, we vary the dark time $\tau_\mathrm{d}$, during which only the tweezer beam is applied, between the end of the evanescent single-stroke loading pulse and the beginning of the excitation pulse. Figure~\ref{fig:lifetime}a shows the raw photon histogram counts as a function of time from the start of the excitation pulse for different values of $\tau_\mathrm{d}$. While the detected signal already begins to decay on timescales shorter than $\tau_\mathrm{d}=1$ ms, signatures of trapped atoms persist for dark times extending up to 1 second.

Figure~\ref{fig:lifetime}b shows the atomic signal as a function of the dark time $\tau_\mathrm{d}$, normalized to the signal at $\tau_\mathrm{d}=0$ for different tweezer wavelengths $\lambda_\mathrm{t}$. The decay spans several orders of magnitude in $\tau_\mathrm{d}$ and, for $\tau_\mathrm{d}\gtrsim 50~\mu$s, is well described by a logarithmic decay of the form $A\log(1+b/\tau_\mathrm{d})$ (solid lines).

Logarithmic relaxation is encountered in a range of physical systems and can arise when the effective loss rate depends exponentially on a broadly distributed parameter, leading to an approximately uniform distribution of decay rates on a logarithmic scale \cite{Amir-2012}. In the present context, a natural candidate is quantum tunneling out of the near-surface trap through a distribution of effective barrier heights, similarly to tunneling in glassy systems \cite{Amir-2012}, which is set by variations in the initial trapping energy and the near-surface potential landscape. In contrast, surface-induced heating mechanisms for neutral atoms near dielectric materials are expected to be weak at these distances \cite{Henkel-1999}. We therefore attribute the dominant loss in the dark to tunneling out of the trap, mainly toward the chip surface, where the barrier is reduced due to the attractive Casimir-Polder potential.

In order to validate this claim, we estimate the tunneling escape rate for different atomic energies using a 1D WKB approximation (see Sec.~\ref{SM_tunneling_rate} of the SM). We find that trapping energies in the range of $1\text{--}5$~MHz below the trap barrier towards the PIC reproduces the observed lifetimes. 
Additionally, this energy distribution is also consistent with the measured number of photons scattered by the atom during the excitation pulse presented in Fig.~\ref{fig:g2}b, assuming recoil heating as the dominant heating mechanism.
This further affirms quantum tunneling as the main loss mechanism limiting the dark lifetime of the atoms in the trap (for a detailed discussion about the trapped atom energies see Sec.~\ref{Loading simulation} of the SM).

Figure~\ref{fig:lifetime}c shows the decay times of the atomic signal to 50\% and 10\% of its initial value as a function of the tweezer wavelength $\lambda_\mathrm{t}$. For $\lambda_\mathrm{t}=835$ and 840~nm, the decay times are on the order of 1~ms for the signal to reach 50\% and on the order of 100~ms to reach 10\%, and are comparable within the experimental uncertainty. The large separation between these two timescales manifests the logarithmic relaxation discussed above. In contrast, for $\lambda_\mathrm{t}=828$~nm the extracted lifetimes are approximately an order of magnitude shorter, which we attribute to the reduced potential barrier of the near-surface trap toward the PIC.
Figure~\ref{fig:lifetime}d shows the dependence of the decay times on the tweezer power for $\lambda_\mathrm{t}=835$~nm. As expected for a tunneling-dominated loss mechanism, increasing the tweezer power increases the trap depth and the height of the potential barrier, resulting in longer atomic lifetimes in the trap.

\section{Discussion and Outlook}
We have demonstrated efficient trapping of a single ultracold rubidium atom on a planar photonic integrated resonator at subwavelength distances of $150\text{--}200$\,nm. Using an evanescent SSL mechanism, atoms are captured near the chip surface in a single irreversible scattering event, without continuous cooling or active feedback.
Despite the extreme proximity to a macroscopic dielectric surface, atoms remain trapped for times extending in rare events up to 1 second, durations that are inaccessible in  non-monolithic structures without continuous cooling\,\cite{Hummer-2019}. This  differs from previous results of trapped atoms near microscopic surfaces, where an exponential decay with lifetimes ranging from a few ms up to $\sim100\,$ms were observed \cite{Will-2021, Goban-2015, Meng-2018, Samutpraphoot-2020}. We attribute the observed large distribution of trapping durations to the broad range of trapped atom energies below the barrier following the SSL process, which in turn gives rise to a large variance in the tunneling rates out of the near-surface trap. This gives good reason to believe that further cooling, either by Raman sideband cooling\,\cite{Meng-2018} or by multiple interactions with a blue-detuned evanescent field\,\cite{Soding-1995, Ovchinnikov-1997, Hammes-2002}, will be effective in significantly extending the trapping lifetimes.

The trapped atom is shown to couple efficiently to the guided resonator mode, exhibiting clear single-emitter signature of photon antibunching, Purcell-enhanced spontaneous emission, and single-atom cooperativity $C\approx 1.5$ at the smallest trap distances. This value is expected to increase considerably already in our next generation of resonators, which exhibit four-fold lower intrinsic loss $\kappa_\mathrm{i}$ thanks to an improved, low-loss ring-bus coupler, an enhanced waveguide-reveal etch recipe, and an optimized cross-section (see Sec. \ref{SM rsonator fit} of the SM).
The loss can be further reduced through advanced fabrication, for example by employing state-of-the-art lithography to minimize sidewall roughness \cite{psiquantum-2025}, expected to yield an approximate factor-of-two improvement. Additional enhancement of the cooperativity can be achieved by incorporating photonic crystal modes to attain strong spatial localization, potentially reaching up to an order of magnitude lower mode volume while maintaining high quality factors \cite{lu-2022}.
Reaching moderate values of $C>30$ is already enough to support deterministic photonic operations with extremely low loss levels and high-fidelity values that are well within the thresholds of existing quantum error-codes \cite{Aqua-2025, Raussendorf2007TopologicalFTCluster}. 
Combined with the inherent compatibility of the SSL scheme with arrays of trapping sites on planar-photonics resonator networks, this establishes the potential of this platform for large scale quantum information processing with photons and atoms.

\section{Methods}

\subsection{Experimental setup}

The optical tweezer operates at wavelengths of $828\text{--}840$ nm with a $1/e^2$ waist radius of 1.5 $\mu$m at the chip position. The tweezer is aligned to the ring-resonator waveguide by imaging the focal spot onto the PIC surface and maximizing the interference fringe pattern that appears when the beam is centered on the waveguide (see Sec.~\ref{SM_position_sensitiviy} of the SM for the sensitivity to the tweezer position).

The excitation beam at 780 nm is combined with the tweezer beam using a dichroic mirror and coupled into the same polarization-maintaining single-mode (PM) fiber. Both beams are launched through a high-NA objective lens toward the chip, with their polarizations aligned parallel to the ring-resonator waveguide. The objective lens (Mitutoyo G Plan Apo 20X) has a numerical aperture of NA=0.28 and a working distance of 30.6\,mm. An additional 852 nm laser beam is sent through the same objective to thermally tune the resonator and lock it to the cycling transition of the rubidium D2 line.

The lensed fiber outputs from the PIC are connected to 99:1 single-mode fiber beam splitters. The 1\% ports are used to inject the resonator locking field and the evanescent single-stroke loading field, and to monitor transmission on an avalanche photodiode. The coupling efficiency from the single-mode fibers into the PIC waveguide is 50\%. The polarization of the resonator locking and loading fields is adjusted to the TM mode using fiber polarization controllers, optimized by minimizing transmission through the PIC.

Photons scattered by trapped atoms into the resonator are collected from both ends of the PIC via the 99\% ports of the fiber beam splitters. At the outputs, the photons pass through waveplates and a polarization beam splitter, a thin-film interference filter, and the first-order diffraction of an acousto-optical modulator (AOM) for temporal gating. The filtered photons are then coupled into PM fibers and detected using the SNSPDs.

\subsection{Experimental sequence}

The experimental cycle begins with a MOT of $N\approx75\times10^6$ $^{87}\text{Rb}$ atoms located approximately 13 mm below the PIC. During the MOT phase, atoms are loaded from a 2D MOT located in a source vacuum cell into a 3D MOT in the main chamber, where the different chambers are separated by a differential pumping tube. After the MOT stage, the atoms are further cooled using polarization gradient cooling (PGC) in an upward moving frame, launching the cloud toward the PIC in an atomic fountain configuration, such that the atoms arrive at the chip with a mean velocity of $\sim 0.3~\mathrm{m/s}$. While these parameters are expected to produce an atomic flux of 0.2\,/ms$/\mu m^2$, with a column density of about about 2/$\mu\mathrm{m}^2$, the tweezer potential which becomes significant near the surface creates a funnel effect and increases the flux by roughly an order of magnitude.

Before the end of the PGC stage, the repump light is turned off, depumping most atoms from the $F=2$ to the $F=1$ ground state. The atomic temperature at the end of the PGC is 30\,$\mu$K. Immediately prior to the evanescent field loading stage, an additional depump pulse resonant with the $F=2\rightarrow F'=2$ transition is applied in order to depump atoms remaining in the $F=2$ ground state to the $F=1$ ground state. 
This depump beam is introduced through the objective lens and is focused at the back focal plane to produce a large beam waist on the ring resonator.

During the MOT stage, the resonator is locked to the cycling transition of the D2 line. Locking is performed by tuning the resonator's temperature with an additional heating beam (wavelength of 852~nm), which minimizes the transmission of the resonator locking beam through the resonator. At the end of the MOT stage, the resonator locking beam is blocked before injection into the PIC.

Upon arrival of the atoms to the PIC, the evanescent single-stroke loading field is applied for 0.5\,ms duration. During the loading pulse, the detection path is gated by switching off the AOMs preceding the SNSPDs to prevent detector saturation from the loading field.
After a variable dark time $\tau_\mathrm{d}$, an excitation pulse is applied to probe the trapped atoms. For most measurements, this loading-excitation sequence is repeated multiple times with an excitation duration of 2~ms and a fixed dark time of 1~$\mu$s. For dark lifetime measurements (Fig.~\ref{fig:lifetime}), the sequence is executed once per atomic cloud launch while varying $\tau_\mathrm{d}$.
For the cooperativity measurements shown in Fig.~\ref{fig:cooperativity}, the excitation beam is further modulated using a semiconductor optical amplifier (SOA) to generate 10 ns-long pulses with 90 ns off-time. 

\subsection{Trap frequencies measurement}

The trap frequencies of a single atom are independently measured using parametric heating (see Sec.~\ref{SM_trap_freq} of the SM for details). We obtain trap frequencies of approximately 650 kHz, 290 kHz, and 60 kHz along the $x$, $y$, and $z$ axes of the trap, respectively, as defined in Fig.~\ref{fig:schematic}.

The trap frequency along the $x$ axis, which is normal to the chip surface, matches the oscillation frequency observed in the normalized intensity autocorrelation function $g^{(2)}(\tau)$ shown in Fig.~\ref{fig:g2}a (inset). This agreement supports the interpretation that the oscillatory correlations arise from atomic motion perpendicular to the resonator surface in the near-surface trap.

\subsection{Trapped-atom excitation resonance}
The excitation spectrum of a trapped atom is measured by recording the photon scattering rate immediately after the evanescent single-stroke loading as a function of the excitation-field detuning $\Delta_\mathrm{e}$ relative to the bare atomic transition. For $\lambda_\mathrm{t}=835$ nm and a tweezer power of 20 mW, the spectral peak is observed at a detuning of $\Delta_\mathrm{e}=85$ MHz relative to a free-space atom, with a half-width at half-maximum of approximately 30 MHz. The lineshape is well described by a Voigt profile. Similar spectra are obtained for $\lambda_\mathrm{t}=840$ nm, whereas for $\lambda_\mathrm{t}=828$ nm the resonance is shifted by an additional $\sim10$ MHz toward the blue.

We note that the excitation detuning that maximizes the total detected atomic signal (integrated over the trapping time until the atom is lost) is $10\text{--}15$ MHz red-detuned from the spectral peak. We attribute this shift to the interplay between excitation-induced heating and near-surface tunneling loss: as the atom scatters photons and gains energy, the tunneling rate increases (see Sec.~\ref{SM_tunneling_rate} of the SM), and optimal signal is obtained at a detuning where the scattering rate and the tunneling loss rate become comparable. All measurements reported in the main text are performed at the detuning that maximizes the detected signal. 

\begin{acknowledgments}
We thank the mechanical engineering, systems engineering, and the control teams at Quantum Source Labs for their collaborative contributions to this project.
We acknowledge the support by the Israel Innovation Authority.
B.D. acknowledges support from the Israel Science Foundation, the U.S.-Israel Binational Science Foundation, the Minerva Foundation, and the United States Army Research Office (Grant No. W911NF-24-1-0392). B.D. holds the Dan Lebas and Roth Sonnewend Professorial Chair of Physics.
\end{acknowledgments}

\bibliography{Sisyphus_paper}


\clearpage

\setcounter{section}{0}
\setcounter{figure}{0}
\setcounter{table}{0}
\setcounter{equation}{0}

\renewcommand{\thesection}{S\arabic{section}}
\renewcommand{\thefigure}{S\arabic{figure}}
\renewcommand{\thetable}{S\arabic{table}}
\renewcommand{\theequation}{S\arabic{equation}}

\begin{center}
\textbf{\large Supplementary Material}
\end{center}


\section{Trap distance from surface versus trap wavelength} \label{trap and g simulations}

The distance of the trap minimum from the PIC surface is controlled by the interference between the incident tweezer field and its partial reflection from the multilayer dielectric stack of the chip. The reflection phase accumulated upon propagation through the SiN and SiO$_2$ layers depends on the tweezer wavelength, thereby shifting the position of the standing-wave nodes and antinodes relative to the surface. By tuning the tweezer wavelength, the trap minimum can be tuned continuously toward or away from the PIC.

The total trapping potential is calculated as the sum of the optical dipole potential generated by the standing-wave tweezer field and the Casimir-Polder potential arising from the dielectric surface. 
We model the Casimir-Polder contribution using the standard short-distance approximation appropriate for atom-surface distances below a few hundred nanometers (see Ref.~\cite{Berroir-2022}).
We calculate the reflected light-field distribution numerically using 3D finite-difference time-domain simulations (Ansys Lumerical). We propagate a Gaussian beam through the PIC layered structure and account for the complex reflection coefficient of the stack. To validate the numerical model, we compare the results of a non-etched SiN thin film to conventional analytical models, finding good agreement.

To evaluate the light-matter interaction, we calculate the mode volume at the atom's position \cite{Chang-2019}. First, the mode area ($A_\mathrm{m}$) of the waveguide cross-section was computed using a 2D finite difference eigenmode solver (Ansys Lumerical). Assuming longitudinal uniformity along the ring resonator, the total mode volume was then approximated by $V_\mathrm{m} \approx 2\pi R A_\mathrm{m}$, where $R$ is the effective radius corresponding to the mode's intensity distribution. From this simulation we extract the decay length $\Lambda = 86\,$nm of the evanescent field of the TM mode at 780\,nm.

\begin{figure*}
\centering
\includegraphics[width=0.75\textwidth]{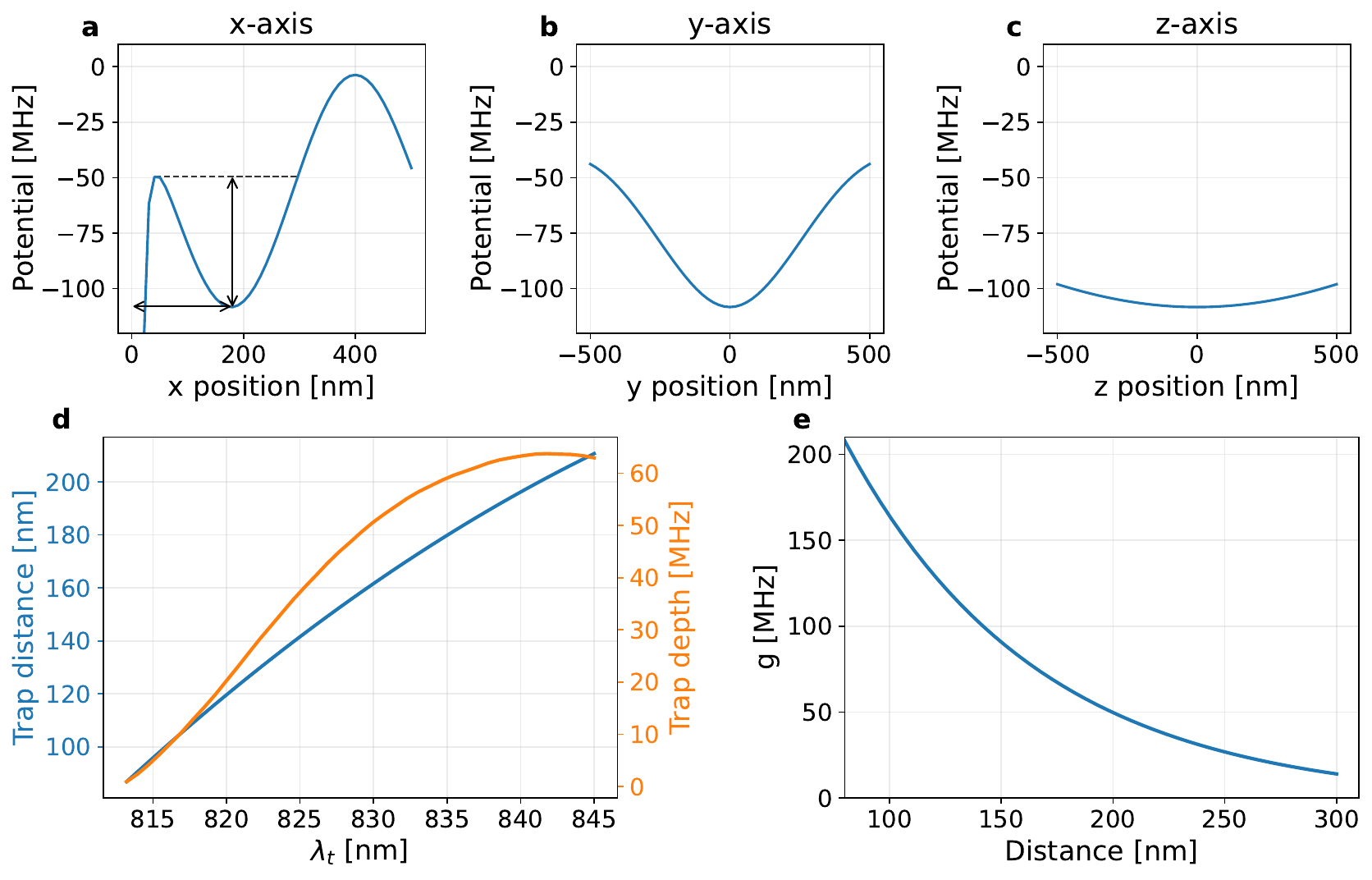}
\caption{\textbf{Simulation of trapping potential and atom-resonator coupling.}
\capa-\capc Calculated trapping potential for an atom in the $F=2$ ground state along the $x$, $y$, and $z$ axes [panels (a), (b), and (c)], corresponding to the distance from the PIC surface, position across the resonator waveguide, and position along the resonator circumference, respectively. The total potential includes the optical tweezer and Casimir-Polder contributions. The horizontal and vertical arrows in (a) define the trap distance (position of the potential minimum relative to the PIC surface) and trap depth, respectively. Parameters: $\lambda_\mathrm{t}=835$ nm, tweezer power 20 mW.
\capd Calculated trap distance (blue) and trap depth (orange) versus tweezer wavelength.
\cape Single-atom coupling rate $g$ versus distance from the resonator surface for the maximally polarized transition $|F=2,m_F=2\rangle \rightarrow |F'=3,m_F'=3\rangle$, evaluated at the waveguide center ($y=0$). The evanescent field of the resonator mode decays exponentially with a decay constant of $\Lambda=86$~nm.  
}
\label{fig:Trap potential}
\end{figure*}

Figures~\ref{fig:Trap potential}a-c show the calculated trapping potential for an atom in the $F=2$ ground state along the $x$, $y$, and $z$ directions for $\lambda_\mathrm{t}=835$ nm and a tweezer power of 20 mW. The trap minimum is located 180 nm from the PIC surface, closer than the $\lambda_\mathrm{t}/4$ distance expected for an ideal standing wave in free space. The potential barrier toward the surface is reduced compared to the free-space case due to both the wavelength-dependent phase shift of the reflected field at short distances and the attractive Casimir-Polder interaction.

Along the $y$ direction, the potential is modified by scattering from the 400 nm-wide waveguide, leading to transverse confinement near the resonator track. Along the $z$ direction, the confinement is dominated by the Gaussian intensity profile of the tweezer beam, which has a waist radius of 1.5 $\mu$m.

Figure~\ref{fig:Trap potential}d shows the calculated trap distance and trap depth as a function of the tweezer wavelength. For $815 \le \lambda_\mathrm{t} \le 845$ nm, decreasing $\lambda_\mathrm{t}$ moves the trap closer to the PIC surface. This simultaneously enhances the atom-resonator coupling strength, as shown in Fig.~\ref{fig:Trap potential}e, and reduces the trap depth due to the increased influence of the Casimir-Polder attraction. The tweezer wavelengths employed in the experiment therefore represent a compromise between maximizing coupling strength and maintaining sufficient trap stability.

\section{Semi-classical simulation for single-stroke loading}\label{Loading simulation}
Here we describe the semi-classical simulation of the SSL mechanism which is used in Fig.\,\ref{fig:param_scan}d. The SSL mechanism is closely related to the physics underlying evanescent-wave surface traps and the gravito-optical surface trap (GOST) introduced by Grimm and co-workers, where spontaneous scattering in a position-dependent optical potential was used to cool an atomic ensemble near a surface\,\cite{Soding-1995, Ovchinnikov-1997, Hammes-2002}.

\textbf{Simulation details.} To gain physical insight into the single-stroke loading mechanism, we perform one-dimensional semi-classical simulations of the atomic dynamics along the direction normal to the chip surface (the $x$ axis). The atomic center-of-mass motion is treated classically. Internal dynamics are reduced to an effective two-level hyperfine manifold corresponding to the $F=1$ and $F=2$ ground states.

The atom is initialized in the $F=1$ state with an initial velocity sampled from the experimentally measured thermal distribution ($T=30\,\mu$K). The total potential experienced by the atom includes contributions from the evanescent loading field, the standing-wave optical tweezer, gravity, and the Casimir-Polder interaction with the chip surface. The optical potentials are state-dependent through the detuning from the relevant hyperfine state. The state-dependent potential due to the evanescent field is given by\,\cite{SteckQuantumAtomOptics}:
\begin{equation}
U(x,F) = \frac{\hbar}{2} \frac{\Delta_F}{1+(\Delta_F/\gamma)^2}\frac{I_0\exp\left(-2x/\Lambda\right)}{I_{\mathrm{sat}}},
\end{equation}
where $\Delta_F$ is the detuning from the bare atomic transition,  $2\gamma=6.06$\,MHz is the natural atomic linewidth, $I_{\mathrm{sat}}=25\,$W/m$^2$ is the saturation intensity, and $I_0$ is the intensity of the evanescent field on the surface of the waveguide.

Photon scattering from the evanescent loading field is incorporated using a Monte-Carlo quantum-jump approach. Over each spatial propagation interval, the position-dependent scattering rate is evaluated from the local intensity and detuning, including light-shift corrections due to the optical trapping potential. The integrated scattering probability over the interval determines stochastically whether a scattering event occurs. The probabilities for spontaneous decay into the $F=1$ and $F=2$ ground states are taken to be 0.75 and 0.25, respectively\,\cite{Soding-1995}. The simulation does not include the repumper field, so scattering induces a single Raman transfer from $F=1$ to $F=2$ (up to off-resonant scattering from the evanescent field back to $F=1$). Once transferred to $F=2$, the atom evolves in the corresponding trap potential for the remainder of the loading pulse. Each scattering event imparts a one-dimensional recoil kick to the atomic velocity.

We simulate each trajectory for the duration of the loading sequence, terminating a trajectory if the atom reaches the surface or is reflected away from the chip. A trajectory is classified as successfully loaded if, following the scattering event, the atom remains within the near-surface trapping region for a significant fraction of the simulation time. The loading probability is obtained by averaging over many stochastic trajectories.

\begin{figure*}
\centering
\includegraphics[width=\textwidth]{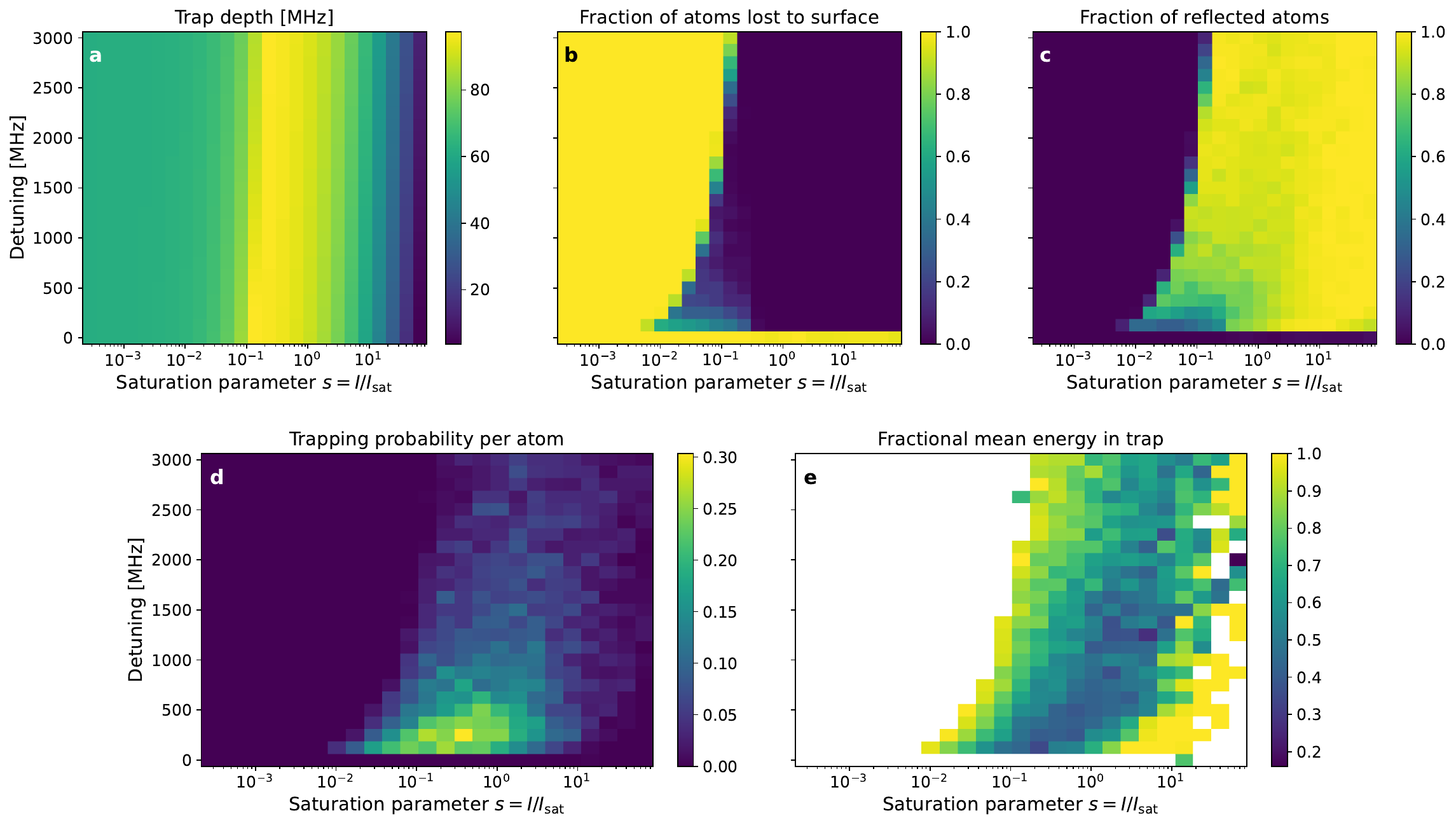}
\caption{\textbf{Semi-classical simulation of the single-scattering loading mechanism.}
Simulation results as a function of evanescent loading-field detuning and saturation parameter $s = I/I_{\mathrm{sat}}$, for a mean incoming atomic velocity of 0.3\,m/s. Each point represents an ensemble of stochastic trajectories simulated using the one-dimensional model described in the text. 
\capa Calculated depth of the near-surface optical potential in the $F=2$ state.
\capb Fraction of atoms that collide with the chip surface during the loading sequence. 
\capc Fraction of atoms that are reflected from the evanescent barrier and escape without being captured. 
\capd Fraction of atoms classified as trapped, defined as trajectories that remain within the trapping region for at least $50\%$ of the simulated duration. 
\cape Mean mechanical energy of atoms that remain in the trap region, normalized to the trap depth $U_0$.
The simulations reveal a well-defined region of optimal loading probability at intermediate detuning and intensity. At low power and detuning, atoms scatter prematurely and are not sufficiently decelerated, leading primarily to surface loss. At high power or large detuning, atoms tend to scatter too close to the surface or after reflection from the evanescent barrier, reducing capture probability. At sufficiently high intensities the evanescent potential significantly distorts the near-surface tweezer potential and suppresses the first standing-wave minimum, eliminating a stable trapping region. }
\label{fig:SSL simulation}
\end{figure*}

\textbf{Simulation results: power-detuning scan.} To visualize the different dynamical outcomes across parameter space, we perform a two-dimensional scan of the evanescent loading-field detuning and saturation parameter $s = I/I_{\mathrm{sat}}$, with the results summarized in Fig.~\ref{fig:SSL simulation}.  Panel~(a) shows the calculated depth of the near-surface trapping potential in the $F=2$ state, illustrating how the evanescent field modifies the standing-wave tweezer potential. At sufficiently large intensities and small detunings, the evanescent-field potential suppresses the local minimum associated with the first
standing-wave antinode even for the $F=2$ state, thereby eliminating the trap. Panels~(b) and~(c) identify the dominant loss channels in the simulation. At low power and large detuning, atoms collide with the chip surface due to a low potential barrier relative to the incoming atomic kinetic energy. At large intensities the atoms are reflected from the evanescent barrier and escape without entering the trapping region. Panel~(d) shows the resulting fraction of trajectories that remain in the trap region for at least $50\%$ of the simulated duration, which we use as an operational definition of successful loading. The simulated loading probability as a function of loading-field detuning and intensity reproduces the main qualitative features of the experimental data, including the existence of an optimal detuning--power regime. Panel~(e) displays the mean motional energy of atoms that remain in the trap region, normalized to the trap depth (see next subsection).

In Fig.\,\ref{fig:param_scan}d we plot the simulation results with the horizontal axis converted to input power at the chip $P_{\mathrm{in}}$.
To this end, we first convert the intensity $I_0$ of the evanescent field on the surface of the waveguide used in the simulation, to circulating power in the waveguide $P_{\mathrm{circ}}$, by using the approximation $I_0=P_{\mathrm{circ}}/A_\mathrm{m}$, where $A_\mathrm{m} = 400 \times 450\,$nm$^2$ is approximately the mode area.
We then calculate the ratio between input chip power and circulating power in the waveguide by accounting for the intensity build-up in the resonator and the chip input loss (50\%). For a critically coupled resonator, the circulating power is given by
\begin{equation}
P_{\mathrm{circ}} = P_{\mathrm{in}} \frac12 \frac{F}{\pi}\frac{1}{1+(\Delta/\kappa)^2}
\end{equation}
where  the factor 1/2 accounts for chip input loss, and the remaining factor accounts for resonator power buildup at the SSL-field detuning ($\Delta\approx6.8\,$GHz).\\

\begin{figure}
\centering
\includegraphics[width=0.5\textwidth]{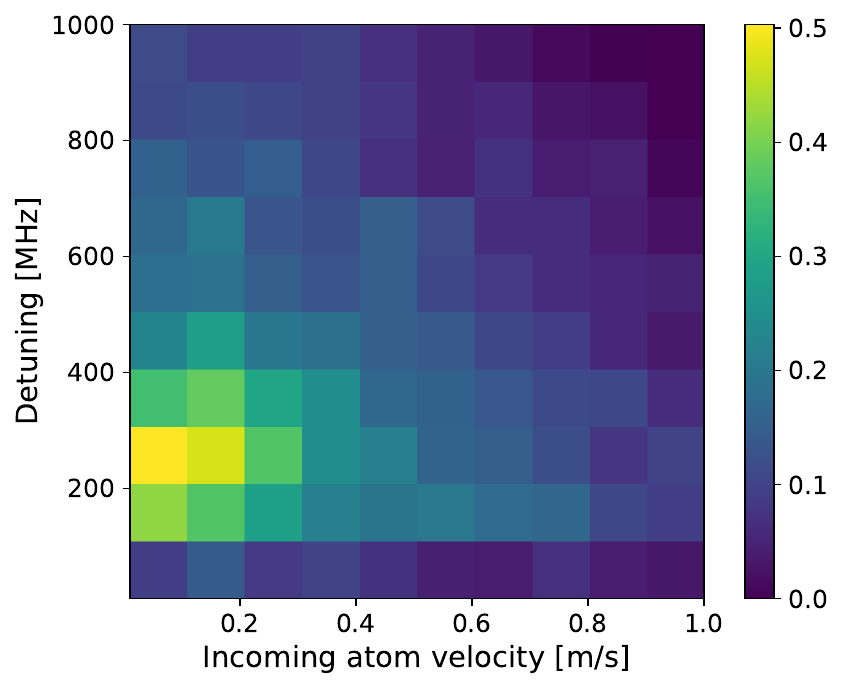}
\caption{\textbf{Simulated loading efficiency as a function of incoming atom velocity and loading-field detuning.}
Trapped atom fraction as a function of incoming atom velocity and evanescent loading-field detuning, calculated for intensity $I/I_{\mathrm{sat}}=2\times10^5$. The loading efficiency is relatively insensitive to the velocity, where atoms with kinetic energies larger than the trap depth can still be efficiently captured. }
\label{fig:Loading vs velocity and detuning}
\end{figure}

\textbf{Sensitivity to incoming atom kinetic energy.} In Fig.\,\ref{fig:Loading vs velocity and detuning}, we test the sensitivity of the SSL mechanism to the initial velocity. The loading efficiency shows a relatively weak dependence on the incoming atomic velocity over a broad range of detunings, in contrast to a 'sudden turn-on' mechanism which requires kinetic energies below the trap depth\,\cite{Will-2021}. This highlights the robustness of the single-scattering capture mechanism, which does not require fine-tuning of the atom's kinetic energy.

\textbf{Simulation limitations.} The model is intentionally minimal and designed to capture the essential physics of irreversible single-scattering capture. Several effects are not included. The full multilevel hyperfine and Zeeman structure of the atom is reduced to an effective two-level system with a fixed Raman branching probability. Excited-state dynamics are not modeled explicitly beyond an effective scattering-rate expression; in particular, tensor and vector light shifts, polarization-dependent optical pumping, and coherent dark-state effects are neglected. Motion is restricted to one dimension, omitting transverse dynamics and spatial variations of the evanescent and tweezer fields across and along the waveguide. The Casimir-Polder interaction is included in a simplified form and does not account for detailed dielectric geometry or position-dependent material response. Technical effects such as stray light, intensity and frequency noise, and other time-dependent experimental imperfections are also neglected.

Despite these simplifications, the simulations reproduce the observed structure of the loading phase diagram and support the interpretation that a single irreversible scattering event is sufficient to capture an atom in the near-surface trap under optimal conditions.

Figure\,\ref{fig:SSL simulation}e shows the mean motional energy of the trapped atoms, normalized to the trap depth (i.e. zero represents the ground state). The optimal simulated mean energy is approximately 0.45 of the trap depth, better than the experimentally inferred value of $\sim0.9$. The reduced trapping performance observed in the experiment could be explained by the fact that most atoms do not arrive exactly on-axis (in the y-direction), and therefore reach shallow regions of the trap.

\section{Resonator characterization and next-generation device performance} \label{SM rsonator fit}

\begin{figure}
\centering
\includegraphics[width=0.5\textwidth]{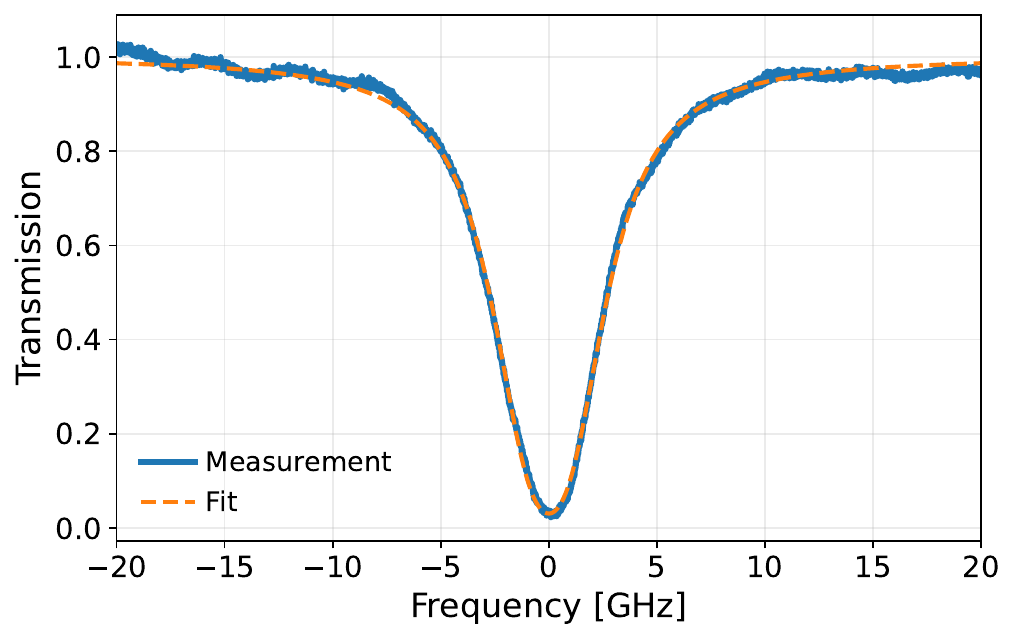}
\caption{\textbf{Resonator transmission spectrum.}
Measured transmission through the photonic integrated resonator as a function of probe-laser detuning from resonance. Solid blue: experimental data. Dashed orange: fit to the standard single-mode resonator transmission model [$T=|t|^2$, where $t$ is given by Eq.~\eqref{eq:transmission_spectrum}]. The fit yields the intrinsic loss rate $\kappa_\mathrm{i}$ and external coupling rate $\kappa_\mathrm{ex}$.}
\label{fig:Resonator spectrum}
\end{figure}

The resonator parameters are extracted by scanning a weak probe laser across the resonance and measuring the transmitted power through the bus waveguide on an avalanche photodiode. Figure~\ref{fig:Resonator spectrum} shows the measured transmission as a function of probe detuning.

The transmission is modeled as $T = |t|^2$, where the complex transmission amplitude $t$ is given by~\cite{Gorodetsky-2000}
\begin{equation} \label{eq:transmission_spectrum}
    t = 1 - \frac{2i\kappa_\mathrm{ex}(i\kappa+\delta)}{(i\kappa+\delta)^2 - h^2}.
\end{equation} 
Here $\delta$ is the detuning of the probe from the cavity resonance, $\kappa = \kappa_\mathrm{ex} + \kappa_\mathrm{i}$ is the total field decay rate ($2\kappa$ is the power decay rate), $\kappa_\mathrm{i}$ is the intrinsic loss rate, $\kappa_\mathrm{ex}$ is the external coupling rate to the bus waveguide, and $h$ denotes the back-scattering coupling rate between the clockwise and counter-clockwise propagating modes.
Fitting the measured spectrum yields
$\kappa_\mathrm{ex} = 1.15 \pm 0.02$ GHz,
$\kappa_\mathrm{i} = 1.16 \pm 0.02$ GHz,
and $h = 1.08 \pm 0.01$ GHz,
implying near critical coupling conditions with on-resonance transmission of 3\%.

Our new QSP4 chip design achieves a significantly higher Q factor (typical four-fold reduction in the intrinsic loss) through several targeted improvements to the resonator geometry and fabrication process. First, the ring-bus coupler was redesigned and optimized based on \cite{Hosseini:10}, using an improved pulley-style coupling geometry to reduce parasitic loss while maintaining efficient coupling to the resonator mode. This minimizes excess scattering and leakage in the coupling region, which is especially important for preserving high intrinsic Q.
Second, the waveguide-reveal process was improved by introducing planarization with chemical-mechanical planarization (CMP) prior to polishing. This provides a smoother and more uniform surface before the wet etch step, enabling better control of the final waveguide roughness and producing a cleaner vacuum-facing interface after reveal. As a result, the exposed optical mode experiences reduced scattering from fabrication-induced surface imperfections.

Finally, the waveguide cross-section was optimized to reduce roughness-induced optical loss while maintaining strong atom–photon interaction. The new design (resonator and bus widths of 720\,nm, 450\,nm, correspondingly, both at 400\,nm  height, with the top $\approx$100\,nm exposed above the silica cladding) balances mode confinement, surface participation, and evanescent-field overlap with the atom. This tradeoff was modeled using the approach described in \cite{Lipson}, allowing us to minimize sensitivity to sidewall roughness without significantly compromising the mode volume, and accordingly the desired atom interaction strength.

\section{Tunneling rates} \label{SM_tunneling_rate}
We calculate the tunneling time of a trapped atom out of the trap for a given atom energy $\Delta E$ relative to the barrier of the potential towards the PIC surface for different tweezer wavelengths. The calculation uses a one-dimensional WKB approximation with the numerically calculated trapping potential in the $x$ direction, as shown in Fig.~\ref{fig:Trap potential}a.

\begin{figure}
\centering
\includegraphics[width=0.5\textwidth]{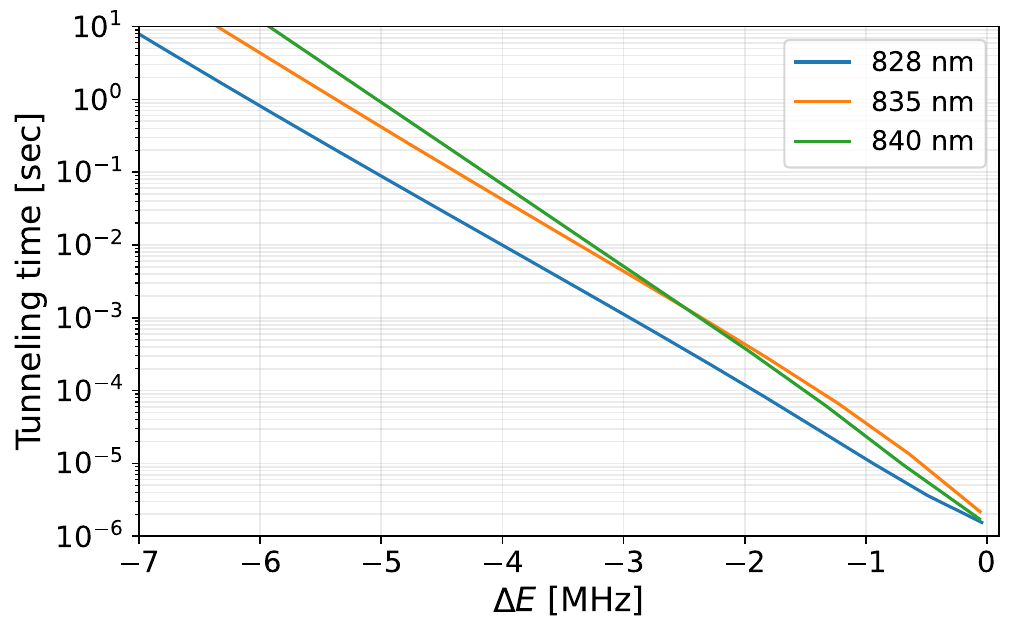}
\caption{\textbf{Tunneling time of a trapped atom.}
Trapped atoms tunneling time out of the trap versus atom energy $\Delta E$ (in frequency units) relative to the potential barrier maximum towards the PIC surface for tweezer wavelengths of 828~nm, 835~nm, and 840~nm and power of 20~mW. A distribution of $-5<\Delta E<-1$\,MHz reproduces the observed lifetimes shown in Fig.~\ref{fig:lifetime} of the main text.}
\label{fig:Tunneling_rates}
\end{figure}

Figure~\ref{fig:Tunneling_rates} shows the tunneling times versus $\Delta E$ in frequency units. The experimentally observed lifetimes, as shown in Fig.~\ref{fig:lifetime} of the main text, are reproduced for $-5<\Delta E<-1$ MHz.

In order to assess if these trapped atoms energies are consistent with our experimental results we consider the number of scattered photons per trapped atom and the spectrum of the trapped atom. As discussed in the main text, a trapped atom scatters $N \sim 100$~photons before escaping the trap. Considering recoil heating as the only heating and loss mechanism $N\approx |\Delta E|/(2E_\mathrm{r})$, where $E_\mathrm{r}$ is the recoil energy induced by a single scattered photon at 780 nm. The detected photon number corresponds to a trapped atom energy of $|\Delta E| \lesssim 1$~MHz, which sets a lower bound for $|\Delta E|$ and is consistent with the estimated trapped atoms energies. Furthermore, the photon number distribution emitted from the trapped atoms (shown in Fig.~\ref{fig:g2}b in the main text) is consistent with the atom energy variation noted above.  

The measured spectrum of the trapped atom has a peak at +85\,MHz away from the bare atom resonance for a tweezer wavelength of 835\,nm and tweezer power of 20\,mW, as discussed in the methods section. At the trap minimum the spectral shift between the ground and excited states relative to the bare transition is 160\,MHz, calculated using the trapping potential simulation. 
We estimate the expected spectral peak shift for different atomic energies by averaging the calculated spectral shift for different trapped atom positions in the 3D classically allowed potential region for its energy. The contribution from each position is weighted by the coupling efficiency of the scattered photon into the resonator $[C(x,y)/(1+C(x,y))]$ and by the time $\Delta t$ the atom spends at the position; $\Delta t$ is estimated by conservation of energy $\Delta t \propto 1/\sqrt{E_\mathrm{k}(x,y,z)}$, where $E_\mathrm{k}(x,y,z)$ is the kinetic energy of the atom at position $(x,y,z)$. For $ -5<\Delta E<-1$\,MHz the expected spectral peak is estimated to be 100-105~MHz. Although the estimated spectrum still differs by $\sim 20\%$ from the measured spectrum, which we attribute to experimental imperfections not considered by the simulation, it indicates that the trapped atoms energies are low relative to the maximal trap depth, in agreement with the atoms energies estimation above.

\section{Details of the trap frequency measurements}\label{SM_trap_freq}

The trap frequencies are measured using parametric heating. After loading a single atom via the evanescent single-stroke mechanism, we sinusoidally modulate the optical tweezer power with a $\pm10\%$ amplitude at a variable frequency. Parametric excitation occurs when the modulation frequency matches twice the trap frequency along one of the principal axes, leading to heating and a reduction in the detected atomic signal~\cite{Jauregui-2001}.

The modulation is applied for 200~$\mu$s following the loading stage. Subsequently, the excitation pulse is applied with constant tweezer power, and the atomic signal is extracted as described in the main text.

\begin{figure}
\centering
\includegraphics[width=0.5\textwidth]{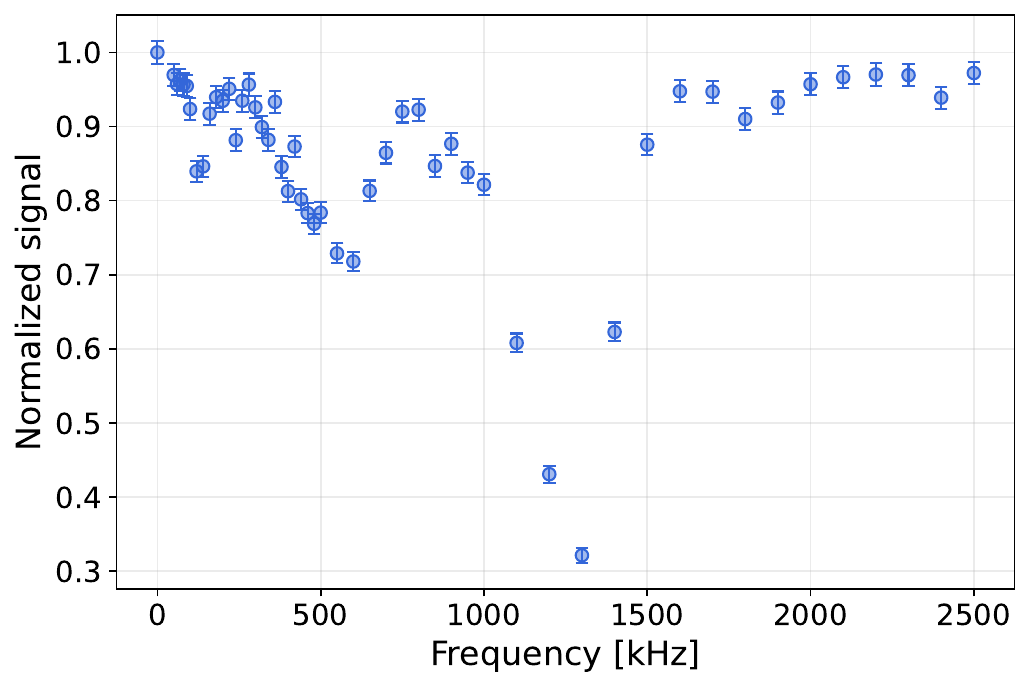}
\caption{\textbf{Trap frequency measurement via parametric heating.}
Normalized atomic signal as a function of the tweezer-power modulation frequency. The signal is normalized to the case without modulation. Three distinct resonances are observed at approximately 1.3 MHz, 580 kHz, and 120 kHz, corresponding to parametric excitation of the trap along the principal axes.
}
\label{fig:Trapfrequency}
\end{figure}

Figure~\ref{fig:Trapfrequency} shows the normalized atomic signal (relative to the case without modulation) as a function of the modulation frequency for $\lambda_\mathrm{t}=835$~nm and a tweezer power of 16 mW. Three distinct resonances are observed at approximately 1.3 MHz, 580 kHz, and 120 kHz. These correspond to trap frequencies of 650 kHz, 290 kHz, and 60 kHz along the $x$, $y$, and $z$ axes of the trap, respectively (as defined in Fig.~\ref{fig:schematic}). 

The highest frequency (650 kHz) corresponds to motion normal to the chip surface ($x$ direction), where the confinement is strongest due to the near-surface standing-wave potential. The intermediate frequency (290 kHz) corresponds to transverse confinement across the waveguide ($y$ direction), and the lowest frequency (60 kHz) reflects the weaker longitudinal confinement along the resonator circumference ($z$ direction) due to the tweezer Gaussian beam.
The $x$-axis frequency agrees with the oscillation frequency observed in the intensity autocorrelation measurements (Fig.~\ref{fig:g2}a).

\section{Tweezer position sensitivity}\label{SM_position_sensitiviy}

\begin{figure}
\centering
\includegraphics[width=0.5\textwidth]{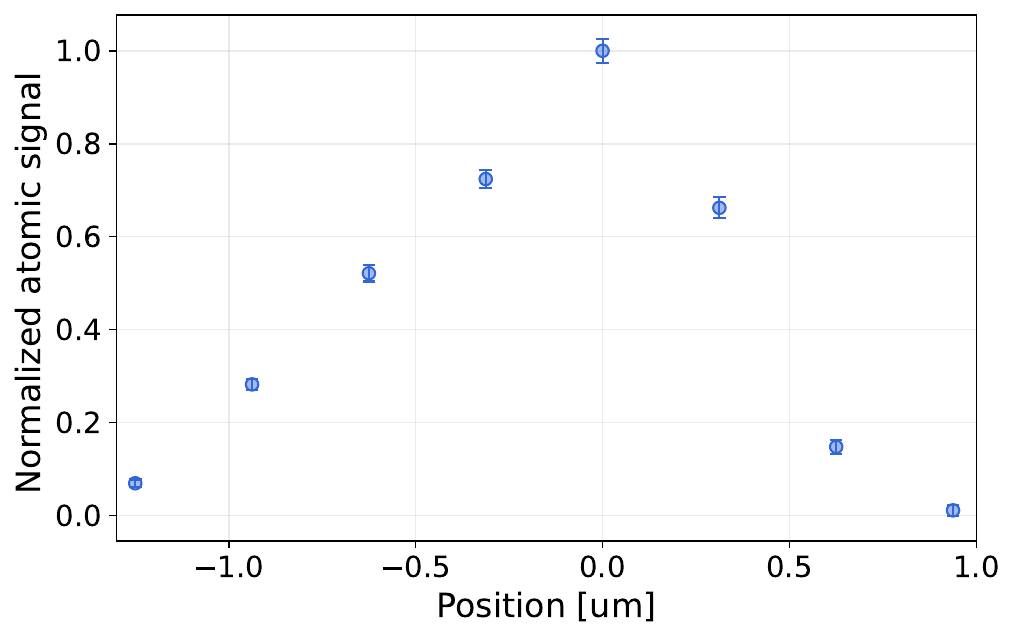}
\caption{\textbf{Tweezer position sensitivity.}
Normalized atomic signal as a function of lateral displacement of the optical tweezer relative to the resonator waveguide. The signal peaks when the tweezer is centered on the waveguide and decreases rapidly for displacements exceeding $\sim0.5~\mu$m, reflecting the spatial localization of the evanescent loading region and the guided resonator mode.}

\label{fig:position scan}
\end{figure}

The trapping performance is sensitive to the lateral alignment of the optical tweezer (waist radius 1.5~$\mu$m) relative to the 450~nm-wide resonator waveguide. To quantify this sensitivity, we scan the tweezer position along the $y$ direction (transverse to the waveguide) and measure the resulting atomic signal, as shown in Fig.~\ref{fig:position scan}.

The scan is performed unidirectionally using a piezo actuator stepper motor (Thorlabs PIA50). The step size is calibrated using the known diameter of the ring resonator. At each position, a short acquisition (15 minutes) is performed to minimize the effect of long-term mechanical drifts.

The atomic signal exhibits a full width at half maximum of approximately 1~$\mu$m, centered on the waveguide position. This width reflects the convolution of the tweezer mode profile with the spatial extent of the evanescent loading region and the guided resonator mode. The slight asymmetry observed in the signal is attributed to deviations from an ideal Gaussian reflection of the tweezer field from the PIC surface.

The reduction in detected signal away from the optimal alignment arises from two effects: a decreased probability for successful loading during the evanescent pulse, and a reduced coupling efficiency of emitted photons into the resonator mode for displaced trap positions.

To probe the effect of alignment on trap stability, we compare measurements performed with dark times of 1~$\mu$s and 1~ms. The optimal tweezer position, defined by the maximum detected atomic signal, also yields the longest observed lifetime, indicating that maximal spatial overlap with the resonator mode improves both loading efficiency and trap stability.

\section{Data analysis} \label{Data analysis SM}

All photon detection events from the SNSPDs are recorded as absolute time tags referenced to a periodically repeated experimental cycle. The data analysis described below uses these time tags to reconstruct time-resolved photon histograms and extract the quantities shown in the main text figures.

For Fig.~2a, photon detection time tags are folded onto the experimental cycle and binned into histograms with 10 $\mu$s resolution. The atomic signal shown in Fig.~2b is obtained by integrating the histogram counts over the interval 0–20 ms, corresponding to the arrival of the atomic cloud at the PIC, and subtracting background counts measured in the same time window after 100 ms, after the atoms near the chip are lost. To extract the trapping probability in Fig.~2c, we count the number of photons detected within the first 500~$\mu$s of each excitation pulse. The analysis is performed on the fourth pulse of the sequence, for which the atomic density at the PIC is maximal. A trapping event is identified when two or more photons are detected within this window. The two-photon threshold suppresses false positive events arising from excitation-field scattering and residual transit signals. The trapping probability is obtained by normalizing the number of such events by the total number of loading attempts. The false-detection probability is estimated from the last 500~$\mu$s of the excitation pulse, where no trapped atoms remain. This yields a false-detection probability of $0.3\pm0.2 \%$, averaged over the full scan of evanescent loading-field detunings and powers.

Figure~3a shows the normalized second-order intensity correlation function $g^{(2)}(\tau)$ between photons detected by the two SNSPDs at the two outputs of the PIC.  
For time-dependent emission, the field is not invariant under time translation, and the second-order coherence therefore needs to be treated as a two-time quantity,
\begin{equation}
g^{(2)}_\mathrm{ab}(t_1,t_2) = 
\frac{N_\mathrm{c} G^{(2)}_\mathrm{ab}(t_1,t_2)}
{G^{(1)}_\mathrm{a}(t_1)G^{(1)}_\mathrm{b}(t_2)},
\end{equation}
rather than as a function of the time-delay alone. Here $G^{(1)}_\mathrm{a(b)}(t)$ denotes the time-dependent single-photon count histogram in detector $\mathrm{a(b)}$, folded onto the experimental cycle, $G^{(2)}_\mathrm{ab}(t_1,t_2)$ is the corresponding two-photon coincidence count histogram, and $N_\mathrm{c}$ is the number of experimental cycles. The histogram in the top panels of Fig.~\ref{fig:param_scan}a is given by $G^{(1)}_\mathrm{a+b}(t)/N_\mathrm{c}$, where $G^{(1)}_\mathrm{a+b}$ the histogram considered from both detectors. 
To obtain the usual one-dimensional correlation function $g^{(2)}(\tau)$, we reparameterize the detection times as $t_1=t$ and $t_2=t+\tau$, defining a delay-resolved correlation $g^{(2)}_\mathrm{ab}(t,\tau)$. Next, we average over the time origin $t$ with the intensity weighting,
\begin{equation} \label{eq: g2 expression}
g^{(2)}_\mathrm{ab}(\tau) =
\frac{N_\mathrm{c} \int dt\, G^{(2)}_\mathrm{ab}(t,t+\tau)}
{\int dt\, G^{(1)}_\mathrm{a}(t)\,G^{(1)}_\mathrm{b}(t+\tau)}.
\end{equation} 
This denominator represents a time-dependent normalization, which accounts for the fact the atom is not constantly emitting light, corresponding in our case for multiple loading pulses and atoms escaping the trap. The integration limits in Eq.~\eqref{eq: g2 expression} are 0--20\,ms, Taking into account the atomic cloud transit and pulse sequence therein. In a discrete implementation, this corresponds to summing all coincidence counts at fixed $\tau$ and normalizing by the sum of the products of the corresponding single-count histograms. In the stationary limit, $G^{(2)}_\mathrm{ab}(t,t+\tau)$ depends only on $\tau$ and $G^{(1)}_\mathrm{a,b}(t)$ become time independent, reducing the above expression to the standard textbook definition of $g^{(2)}(\tau)$.

Correlations computed over all trapping attempts exhibit an antibunching feature, but the dip does not reach sub-Poissonian values ($g^{(2)}(0)\approx 1$) due to the near-Poissonian atom loading statistics. 
We therefore condition the analysis on successful trapping events as defined above. Photon time tags are taken from the first $500~\mu$s of the excitation pulse and from the first eight pulses during which atoms arrive at the PIC (Fig.~2a), and the correlation function is normalized using only pulses in which a trapped atom was detected. The error bars are calculated assuming Poissonian photon counting statistics. For the inset of Fig.~3a, a $77\,$ns-wide sliding-window average of $g^{(2)}(\tau)$ is applied for $\tau>50$~ns to highlight the oscillations associated with atomic motion in the trap.

Figure~3b shows the probability $P(N)$ to detect $N$ photons within the first $500~\mu$s of the excitation pulse. The statistics are compiled from the first eight excitation pulses of the sequence, during which atoms arrive at the PIC (Fig.~2a). Consistent with the trapping criterion defined above, the distribution is analyzed for $N \ge 2$ and is well described by an exponential decay $P(N)=A\exp(-N/N_0)$.

Figure~4a shows the photon-count histogram obtained by folding the SNSPD time tags onto the $100$~ns excitation-pulse cycle. The total signal is calculated from photon counts detected during the first $20$~ms of the experimental sequence (eight loading–excitation cycles), when atoms transit near the PIC. The background signal associated with leakage of the excitation pulse is obtained from the same time interval after $100$~ms, when no atoms remain near the chip. The atomic fluorescence signal is obtained by subtracting this background contribution from the total signal. The decay is fitted to an exponential function with a constant background, $A\exp(-t/\tau_\mathrm{e})+B$, for times $t>10$~ns, after the excitation pulse has fully decayed.

As described in the main text and shown in Fig.~4b, the cooperativity is extracted from the Purcell-enhanced lifetime using $C=\Gamma/\gamma-1$, and the atom-cavity coupling rate is obtained from $C=g^2/(\kappa\gamma)$. The error bars are derived from propagation of the uncertainty of the exponential decay fit. 

The shaded region in Fig.~4b represents the range of coupling strengths expected from the unknown Zeeman-state distribution of the trapped atoms. The upper bound corresponds to the maximally polarized cycling transition $|F=2,m_F=2\rangle\rightarrow|F'=3,m'_F=3\rangle$, which yields the largest possible coupling to the resonator mode.

The lower bound is obtained by assuming an equal population of the excited-state Zeeman sublevels and defining an effective rms dipole moment $\mu_{\mathrm{eff}}$ for coupling to the $\sigma^\pm$ resonator modes,
\begin{equation}
    \mu_{\mathrm{eff}}=
    \sqrt{\frac{1}{N}\sum_{m'_F}\sum_{q=\pm1}\mu^2_{|F'=3,m'_F\rangle\rightarrow|F=2,m_F=m'_F+q\rangle}},
\end{equation}
where $q=\pm1$ corresponds to $\sigma^\pm$ transitions and $N=7$ is the number of $m'_F$ sublevels included in the average. This yields $\mu_{\mathrm{eff}}\simeq 0.816\,\mu_{m'_F=3\rightarrow m_F=2}$, and the corresponding lower bound of $g$ is obtained by scaling the maximal coupling strength by the ratio $\mu_{\mathrm{eff}}/\mu_{m'_F=3\rightarrow m_F=2}$.

Figure~5a shows histograms of the photon detection events for different dark times $\tau_\mathrm{d}$ between the end of the loading pulse and the start of the excitation pulse. The SNSPD time tags are folded onto the experimental cycle and binned with a temporal resolution of $100~\mu$s. 
The histograms illustrate the decay of the trapped-atom fluorescence signal as the dark time $\tau_\mathrm{d}$ is increased.

Figure~5b shows the normalized atomic signal as a function of the dark time $\tau_\mathrm{d}$. The signal is obtained by integrating the photon counts within the first $500~\mu$s of the excitation pulse and subtracting the background counts measured after $100$~ms of excitation, when trapped atoms have already been lost from the trap. The resulting signal is normalized to the signal at $\tau_\mathrm{d}=0$. The integration window is limited to the first $500~\mu$s of the excitation pulse in order to avoid contributions from residual transit events. The error bars are calculated assuming Poissonian photon counting statistics.

The decay times reported in Figs.~5c,d are extracted from the normalized atomic signal shown in Fig.~5b. For each dataset, the signal as a function of $\tau_\mathrm{d}$ is fitted with a smoothing spline, with the smoothing parameter determined by a generalized cross-validation criterion. The decay times are then defined as the values of $\tau_\mathrm{d}$ at which the spline reaches $50\%$ and $10\%$ of the signal at $\tau_\mathrm{d}=0$.

Figure~5d shows the results from a single measurement. The uncertainties are estimated by adding random noise to the measured signal using a normal distribution with standard deviation given by the Poissonian counting statistics. The spline fit is recomputed for each realization, and the decay times are extracted as described above. The error bars correspond to the standard deviation of the resulting distribution of decay times.

Figure~5c includes multiple measurements for $\lambda_\mathrm{t}=835$~nm performed over three months, in addition to single measurements for $\lambda_\mathrm{t}=828$~nm and $\lambda_\mathrm{t}=840$~nm. The decay times extracted from individual runs exhibit a broad multiplicative spread and are therefore analyzed in logarithmic space. For a set of measured decay times $\tau_i$, we define $x_i=\log\tau_i$ and compute the mean $\mu_x$ and standard deviation $\sigma_x$ of the $x_i$ distribution. In this representation $\sigma_x$ corresponds to the relative (multiplicative) uncertainty of the decay times. The central value reported in Fig.~5c is the geometric mean $\tau_{\mathrm{geo}}=\exp(\mu_x)$, and the asymmetric $1\sigma$ uncertainties are obtained by propagating $\sigma_x$ back to linear space. The relative uncertainty $\sigma_x$ extracted from the repeated measurements at $\lambda_\mathrm{t}=835$~nm is also used to estimate the uncertainties of the single measurements at $\lambda_\mathrm{t}=828$~nm and $\lambda_\mathrm{t}=840$~nm.

\end{document}